\documentclass[10pt,a4paper,twocolumn]{article}
\usepackage[T1]{fontenc}
\usepackage[utf8]{inputenc}
\usepackage{lmodern,microtype}
\input{glyphtounicode}\pdfgentounicode=1
\usepackage[margin=18mm,columnsep=6mm]{geometry}
\usepackage{amsmath,amssymb,graphicx,booktabs,tabularx,array,longtable}
\usepackage[dvipsnames]{xcolor}
\usepackage[font=small,labelfont=bf]{caption}
\usepackage[numbers,sort&compress]{natbib}
\usepackage{xurl}
\usepackage[colorlinks=true,allcolors=MidnightBlue]{hyperref}
\usepackage{fancyhdr}
\usepackage{xspace}
\usepackage{tikz}\usetikzlibrary{positioning}
\usepackage[most]{tcolorbox}
\newcommand{\nApportApplication}{19\xspace}
\newcommand{\nApportNew}{28\xspace}
\newcommand{\nApportVariant}{102\xspace}
\newcommand{\nAssessed}{298\xspace}
\newcommand{\nAxesAmbiguous}{41\xspace}
\newcommand{\nCitExcluded}{588\xspace}
\newcommand{\nCitKeptRescreen}{10\xspace}
\newcommand{\nCitNoAbstract}{136\xspace}
\newcommand{\nCitRetained}{105\xspace}
\newcommand{\nCitScrExclApplication}{13\xspace}
\newcommand{\nCitScrExclHierarchie}{10\xspace}
\newcommand{\nCitScrExclHorsSujet}{35\xspace}
\newcommand{\nCitScrExclNonSpatial}{46\xspace}
\newcommand{\nCitScrExclObjetsSepares}{0\xspace}
\newcommand{\nCitScrExclPasUnArticle}{0\xspace}
\newcommand{\nCitScrExclPatchsEntrainement}{0\xspace}
\newcommand{\nCitScrExclRegionsPendant}{86\xspace}
\newcommand{\nCitScrExclRetracte}{0\xspace}
\newcommand{\nCitScrExclRevueGenerale}{15\xspace}
\newcommand{\nCitScrExclSansPartition}{383\xspace}
\newcommand{\nCitScrExclUneDimension}{0\xspace}
\newcommand{\nCitScreened}{693\xspace}
\newcommand{\nClaims}{512\xspace}
\newcommand{\nClaimsReworded}{52\xspace}
\newcommand{\nClaimsVerified}{460\xspace}
\newcommand{\nDimN}{1\xspace}
\newcommand{\nDimThree}{6\xspace}
\newcommand{\nDimTwo}{142\xspace}
\newcommand{\nDisagreeScreening}{21\xspace}
\newcommand{\nDuplicates}{1{,}030\xspace}
\newcommand{\nExclApplication}{3\xspace}
\newcommand{\nExclHierarchie}{1\xspace}
\newcommand{\nExclHorsSujet}{5\xspace}
\newcommand{\nExclNonSpatial}{17\xspace}
\newcommand{\nExclObjetsSepares}{5\xspace}
\newcommand{\nExclPasUnArticle}{0\xspace}
\newcommand{\nExclPatchsEntrainement}{1\xspace}
\newcommand{\nExclRegionsPendant}{14\xspace}
\newcommand{\nExclRetracte}{1\xspace}
\newcommand{\nExclRevueGenerale}{0\xspace}
\newcommand{\nExclSansPartition}{71\xspace}
\newcommand{\nExclUneDimension}{3\xspace}
\newcommand{\nExcludedFullText}{121\xspace}
\newcommand{\nExcludedScreening}{572\xspace}
\newcommand{\nFamAcquisition}{11\xspace}
\newcommand{\nFamAdaptive}{4\xspace}
\newcommand{\nFamFixed}{55\xspace}
\newcommand{\nFamManual}{1\xspace}
\newcommand{\nFamOtherFamily}{1\xspace}
\newcommand{\nFamRegions}{76\xspace}
\newcommand{\nFamTwoLevel}{1\xspace}
\newcommand{\nFullTextAgree}{57\xspace}
\newcommand{\nFullTextSample}{58\xspace}
\newcommand{\nGTandOVM}{25\xspace}
\newcommand{\nGroundTruth}{94\xspace}
\newcommand{\nGroups}{17\xspace}
\newcommand{\nIEEE}{186\xspace}
\newcommand{\nKeptRescreen}{5\xspace}
\newcommand{\nMergeOrderPairs}{2\xspace}
\newcommand{\nModHolography}{8\xspace}
\newcommand{\nModMRI}{15\xspace}
\newcommand{\nModOptical}{42\xspace}
\newcommand{\nModOtherModality}{4\xspace}
\newcommand{\nModProfilometry}{16\xspace}
\newcommand{\nModSAR}{64\xspace}
\newcommand{\nNotRetrieved}{5\xspace}
\newcommand{\nOVMNumber}{25\xspace}
\newcommand{\nOVMNumberTruth}{11\xspace}
\newcommand{\nOVMTimeOnly}{5\xspace}
\newcommand{\nOVMTimeOrVisual}{18\xspace}
\newcommand{\nOVMVisual}{13\xspace}
\newcommand{\nOneVsMany}{43\xspace}
\newcommand{\nOpenAlex}{657\xspace}
\newcommand{\nOrgDuring}{14\xspace}
\newcommand{\nOrgJoint}{23\xspace}
\newcommand{\nOrgNotApplicable}{1\xspace}
\newcommand{\nOrgNotReported}{25\xspace}
\newcommand{\nOrgOneAtATime}{72\xspace}
\newcommand{\nOrgReference}{10\xspace}
\newcommand{\nOrgSingleIntegration}{4\xspace}
\newcommand{\nPartialReading}{2\xspace}
\newcommand{\nPeriodFour}{42\xspace}
\newcommand{\nPeriodOne}{19\xspace}
\newcommand{\nPeriodThree}{54\xspace}
\newcommand{\nPeriodTwo}{34\xspace}
\newcommand{\nRealData}{138\xspace}
\newcommand{\nRecordsRaw}{1{,}800\xspace}
\newcommand{\nReportsIncluded}{177\xspace}
\newcommand{\nRetainedRecords}{198\xspace}
\newcommand{\nRetainedWorks}{188\xspace}
\newcommand{\nScopus}{547\xspace}
\newcommand{\nScrExclApplication}{70\xspace}
\newcommand{\nScrExclHierarchie}{4\xspace}
\newcommand{\nScrExclHorsSujet}{192\xspace}
\newcommand{\nScrExclNonSpatial}{28\xspace}
\newcommand{\nScrExclObjetsSepares}{0\xspace}
\newcommand{\nScrExclPasUnArticle}{0\xspace}
\newcommand{\nScrExclPatchsEntrainement}{21\xspace}
\newcommand{\nScrExclRegionsPendant}{6\xspace}
\newcommand{\nScrExclRetracte}{0\xspace}
\newcommand{\nScrExclRevueGenerale}{23\xspace}
\newcommand{\nScrExclSansPartition}{209\xspace}
\newcommand{\nScrExclUneDimension}{19\xspace}
\newcommand{\nScreenSample}{154\xspace}
\newcommand{\nScreened}{770\xspace}
\newcommand{\nSeeds}{76\xspace}
\newcommand{\nSeedsExcluded}{5\xspace}
\newcommand{\nSensMaxShift}{4\xspace}
\newcommand{\nSensOtherAlt}{8\xspace}
\newcommand{\nSought}{303\xspace}
\newcommand{\nSoughtCit}{105\xspace}
\newcommand{\nSoughtDB}{188\xspace}
\newcommand{\nSoughtOther}{10\xspace}
\newcommand{\nStudiesIncluded}{149\xspace}
\newcommand{\nTabEstimators}{63\xspace}
\newcommand{\nVarInteger}{80\xspace}
\newcommand{\nVarNone}{21\xspace}
\newcommand{\nVarNotReported}{17\xspace}
\newcommand{\nVarOther}{15\xspace}
\newcommand{\nVarReal}{16\xspace}
\newcommand{\nWoS}{410\xspace}
\newcommand{\yearFirst}{1987\xspace}
\newcommand{\yearLast}{2026\xspace}

\newcommand{\nGroupNR}{12\xspace}

\newcommand{\W}{\mathcal W}
\newcommand{\phraseQueriesAgent}{
A commercial AI agent helped explore and refine the queries.}
\hypersetup{pdftitle={Partitioning and Merging in Phase Unwrapping: A Systematic Review of Methods and Evidence},pdfauthor={Antoine Moevus, Max Mignotte},pdfkeywords={phase unwrapping, partitioning, tiling, block-wise processing, region segmentation, region merging, phase stitching, subaperture stitching, mosaicking, domain decomposition, divide and conquer, quadtree, parallel computing, HPC, InSAR, MRI, digital holography, fringe projection profilometry, optical metrology, systematic review}}

\hypersetup{pdfkeywords={Tiling, Block-wise processing, Region-based methods, Phase stitching, Parallel phase unwrapping, Large-scale phase unwrapping, Congruence}}
\title{Partitioning and Merging in Phase Unwrapping:\\ A Systematic Review of Methods and Evidence}
\author{Antoine Moevus and Max Mignotte\\\small D\'epartement d'informatique et de recherche op\'erationnelle\\\small Universit\'e de Montr\'eal, Canada}
\date{November 2026}
\begin{document}
\maketitle
\begin{abstract}
\setlength{\parindent}{0pt}\setlength{\parskip}{0.5\baselineskip}\noindent
Phase unwrapping underlies radar interferometry, magnetic resonance imaging and optical metrology.
When an image is too large or too difficult to unwrap in one piece, many methods cut it into units, unwrap each unit, and merge the results.
To our knowledge, no systematic review had brought these partitioned methods together.

This review gathers them, for two-dimensional images and three-dimensional volumes.
It describes how the units are built: fixed tiles, adaptive tiles, regions computed from the data, or pieces imposed by the acquisition, such as subapertures, slices and scenes.
It then classifies how the units are merged, in four classes: after each unit is unwrapped, through one integer per jump-free region, during the solve, or through an external reference.

Three findings stand out: building the units and merging them are two separate decisions; the merging rule decides whether the merged phase stays congruent with the measured one; and most tests change both decisions at once, so few results show which one helps.
We state what a study can report to support a claim on memory, speed or accuracy.
Our aim is to document the methods, not to rank them.

Following PRISMA 2020, four bibliographic databases (OpenAlex, IEEE Xplore, Web of Science and Scopus), searched on 1 October 2026, returned \nRecordsRaw records.
After removing duplicates, screening titles and abstracts, adding a citation search and reading \nAssessed full texts, we included \nStudiesIncluded studies, published from \yearFirst to \yearLast.
The review was not registered and received no specific funding.
\end{abstract}

\section{Introduction}
\label{sec:intro}
Many imaging techniques measure a phase only modulo $2\pi$: each value is wrapped into an interval of width $2\pi$.
Phase unwrapping recovers the continuous phase from these wrapped values.
This step is part of synthetic aperture radar interferometry (InSAR), magnetic resonance imaging (MRI) and optical metrology.

As these images grow larger, unwrapping them in one piece strains the available memory and computing time.
Many methods therefore cut the image into smaller tiles or regions, which this review calls units.
Each unit is unwrapped separately, and the local solutions are then merged into one phase map.
We call these methods partitioned.

Memory and time are not the only reasons to partition an image, and the other reasons fall into two kinds.
The first kind comes from the computation.
Tiles can bring a large interferogram, the phase image of InSAR, within reach of a network-flow solver~\citep{chen2002phase}.
Tiles can also be assigned to parallel processes~\citep{huang2015parallel,onat2026parallel}.
Finally, a neural network with a fixed input size imposes its own tiles~\citep{zhao2021phase}.

The second kind of reason concerns the quality of the result.
A small tile can carry a simple local model, such as a polynomial fitted to its phase~\citep{friedlander1996model,antonopoulos2015tilebased}.
Partitioning can also confine errors.
When the phase is integrated along paths, unit boundaries limit how far an error spreads along a path: errors mostly stay inside their tile~\citep{antonopoulos2015tilebased}.
In the same way, a cut along a physical discontinuity keeps unwrapping from linking points that are close but physically separate~\citep{song2024toward}.
With least squares, the same effect comes from the size of the solve.
A global least-squares solver spreads the mismatch of inconsistent differences over the whole image, whereas a solve on a small tile involves only the pixels of that tile.
For this reason, Moevus and Mignotte use tiles for least-squares unwrapping~\citep{moevus2026tilebased}.

\begin{figure*}[tbp]
\centering\includegraphics[width=0.86\linewidth]{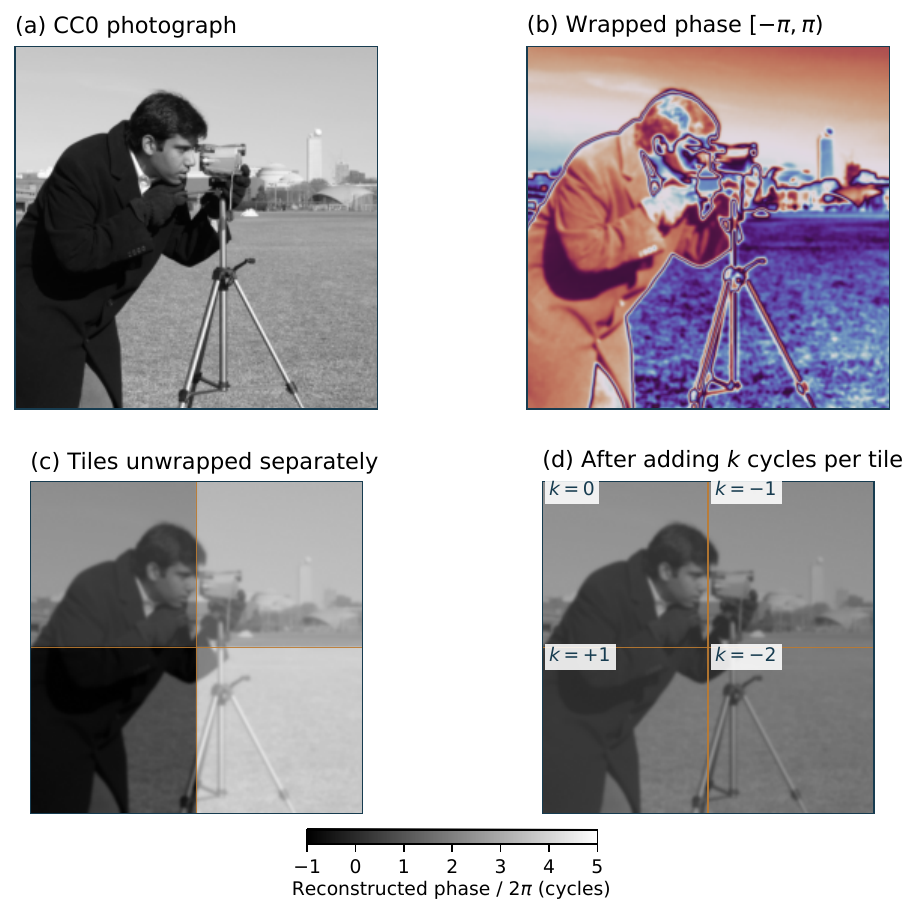}
\caption{\textbf{Four tiles, each correct up to a whole number of cycles, form a wrong mosaic.}
(a)~A photograph used as a spatial template (the CC0 \emph{camera} image of scikit-image).
(b)~The wrapped synthetic phase $\W(6\pi\,G_2*I)$, where $I$ is the photograph scaled to $[0,1]$, $G_2$ is a Gaussian kernel of standard deviation two pixels, $*$ denotes convolution and $\W$ wraps its argument into $[-\pi,\pi)$.
(c)~Four $256\times256$ tiles unwrapped separately, each shifted by a whole number of cycles chosen for this illustration.
(d)~The same tiles after adding to each tile the integer number of cycles $k$ that corrects it.
Both mosaics wrap exactly to~(b).}
\label{fig:photo}
\end{figure*}

Whatever the reason for the cut, merging the local solutions raises a problem of its own.
When the local solver keeps each result consistent with the wrapped data, each unit is known only up to an unknown number of cycles, where one cycle is $2\pi$.
Two neighboring units can therefore both be correct and still disagree by a whole number of cycles.
Figure~\ref{fig:photo} shows this ambiguity on an image cut into four tiles: every tile in panel~(c) wraps to the observed phase, yet the mosaic is wrong.
Because the wrapped data alone do not say which tile should move, the merging rule must supply that information.

This merging problem arises in every modality, yet, to our knowledge, no existing review gathers the partitioned methods in a systematic way.
The book of Ghiglia and Pritt surveys two-dimensional unwrapping algorithms in general~\citep{ghiglia1998two}.
Other reviews cover one modality, such as InSAR~\citep{yu2019phase}, or one family of methods, such as deep learning~\citep{wang2022deep}.

This review fills that gap.
It organizes the partitioned methods around two decisions: how a method builds its units, and how it merges their solutions.
It describes what the studies report, without re-running or re-implementing any method.
Its contributions are three.
First, it sorts \nStudiesIncluded studies across imaging modalities into \nGroups groupings by mechanism.
It places all but \nGroupNR of these studies into four classes, according to where the relation between units is decided; the other \nGroupNR describe their partition but not their merging.
Second, it describes the methods by their two decisions and by what each unit returns.
Third, it shows what the reported comparisons establish, and where evidence is missing.

The rest of the review follows this order.
Section~\ref{sec:background} defines the notions used throughout.
Section~\ref{sec:methods} gives the methods of the review, and Section~\ref{sec:included} the studies it includes.
Section~\ref{sec:taxonomy} then presents the taxonomy, the main result, and Section~\ref{sec:evidence} examines the published evaluations.
Finally, Section~\ref{sec:discussion} discusses the findings and their limits.

\section{Definitions: congruence, units and merging}
\label{sec:background}
This section defines the notions used in the rest of the review.

\paragraph{Congruence.}
Let $\phi$ be the unknown continuous phase, and let the measured phase be $\psi=\W(\phi)$, where $\W$ wraps any value into $[-\pi,\pi)$.
An unwrapped result $u$ is \emph{congruent} when $\W(u)=\psi$ at every pixel.
Equivalently, $u=\psi+2\pi k$, where $k$ is a field of integers, one per pixel.
The integer $k$ of a pixel is its number of cycles.
Congruence can be checked without knowing the true phase.
A rewrapping residual measures it: at every pixel, this residual compares $\W(u)$ with $\psi$ as angles~\citep{moevus2026tilebased}.
In practice, the measurement also contains noise: $\psi=\W(\phi+n)$, where $n$ is the noise.
The noise-free phase $\phi$ is then generally not congruent with $\psi$, so congruence and accuracy are different things.

\paragraph{Jumps, wrapped differences and residues.}
Congruence does not settle $k$ either: every choice of $k$ gives a congruent field.
The choice must therefore come from spatial assumptions, additional measurements or the quantity that the algorithm minimizes, called its objective.
To make this choice, methods read the trace that the wrapping leaves in $\psi$: a \emph{jump}, also called a phase wrap, which is a change of more than $\pi$ between two neighboring pixels.
Where the phase varies slowly, this change is close to $2\pi$.
A band of the image between two jumps is a \emph{fringe}, and in optics its number of cycles is its fringe order.
In addition, the true phase can contain physical discontinuities, for example at the edge of an object.
Unwrapping must remove the jumps but keep these discontinuities.
For both tasks, the basic tool is the wrapped difference of two neighboring pixels: their difference in $\psi$, wrapped into $[-\pi,\pi)$.
This wrapped difference equals the difference before wrapping whenever the phase before wrapping changes by less than $\pi$ between the two pixels.
Integration sums these differences along a path to rebuild the phase.
This sum can fail: a \emph{residue} is a loop of $2\times2$ pixels around which the wrapped differences sum to $+2\pi$ or $-2\pi$ instead of zero~\citep{goldstein1988satellite}.
The sign of this sum is the \emph{polarity} of the residue.
Around a residue, the result of integration depends on the path.

\paragraph{Classical methods.}
Classical methods differ in how they handle these residues and choose $k$, and they fall into three families.
First, path-following methods integrate pixel by pixel.
Among them, branch-cut methods join residues of opposite polarity by lines called branch cuts, so that the polarities on each cut sum to zero~\citep{goldstein1988satellite}.
They then integrate only along paths that cross no cut.
Quality-guided methods, which also follow a path, unwrap the most reliable pixels first, according to a quality map.
Both kinds can also skip pixels: a mask marks the pixels that are not unwrapped, such as pixels without valid data.
Second, least-squares and minimum-$L^p$-norm methods do not follow a path: they choose the continuous field that minimizes one objective~\citep{ghiglia1994robust,ghiglia1996minimum}.
This objective sums, over all pairs of neighboring pixels, the departure of the unwrapped difference from the wrapped difference, raised to the power $p$.
With $p=2$, the method is least squares, which, without weights, amounts to solving a Poisson equation with Neumann boundary conditions.
With $p=1$, the objective adds up the sizes of the departures, and with $p=0$, it counts them.
Third, network-flow and discontinuity-minimizing methods work on integers~\citep{flynn1997two,costantini1998novel,chen2001twodimensional}.
In a network-flow method, a minimum-cost flow solver chooses an integer number of cycles to add to each wrapped difference, so that all residues are removed at the least total cost.
The integers $k$ then follow by integration.
When the cost is proportional to the number of cycles added, this is the $p=1$ objective, with weights, restricted to whole numbers of cycles~\citep{costantini1998novel}.
A discontinuity-minimizing method chooses $k$ directly, so that the jumps left between neighboring pixels of the unwrapped phase have the smallest weighted sum of sizes, counted in cycles.
Its author calls these remaining jumps discontinuities~\citep{flynn1997two}.
Any of these methods can run inside the units of a partitioned method; it is then called the \emph{local solver}, and one run of it on one unit is a \emph{local solve}.

\paragraph{Units.}
A \emph{unit} is a part of the image that is unwrapped or labeled as one piece, then placed as a whole by the merging, for example by one offset.
It can be a fixed tile, a region that receives one integer, the area covered by a local model, or a tile drawn around a group of residues.
The studies call their tiles blocks, patches, sub-images or zones, and this review keeps the word of each study.

\paragraph{Merging.}
\emph{Merging} turns the outputs of the units into one phase field.
Two neighboring units meet along a \emph{seam}, the boundary or the overlap that they share.
Suppose that unit $i$ returns a phase map $\widehat u_i$ on its set of pixels $\Omega_i$.
A common merging model is then
\begin{equation}
 u_i(x)=\widehat u_i(x)+c_i,\qquad x\in\Omega_i,
 \label{eq:offset}
\end{equation}
where the offset $c_i$ is one constant added to every pixel of the unit.
A \emph{seam estimator} is the rule that turns the differences measured along a seam into one offset.
If $\widehat u_i$ is congruent, adding an integer multiple of $2\pi$ keeps it congruent, whereas a real-valued offset generally does not.
A continuous least-squares estimate $\widehat u_i$ is generally not congruent~\citep{chen2001twodimensional}; no integer offset can then make it congruent, because adding a multiple of $2\pi$ leaves its rewrapped values unchanged.
Other units return integer labels or branch cuts instead of a phase map, so Equation~\eqref{eq:offset} does not describe them.

\begin{figure*}[tbp]
\centering\includegraphics[width=\linewidth]{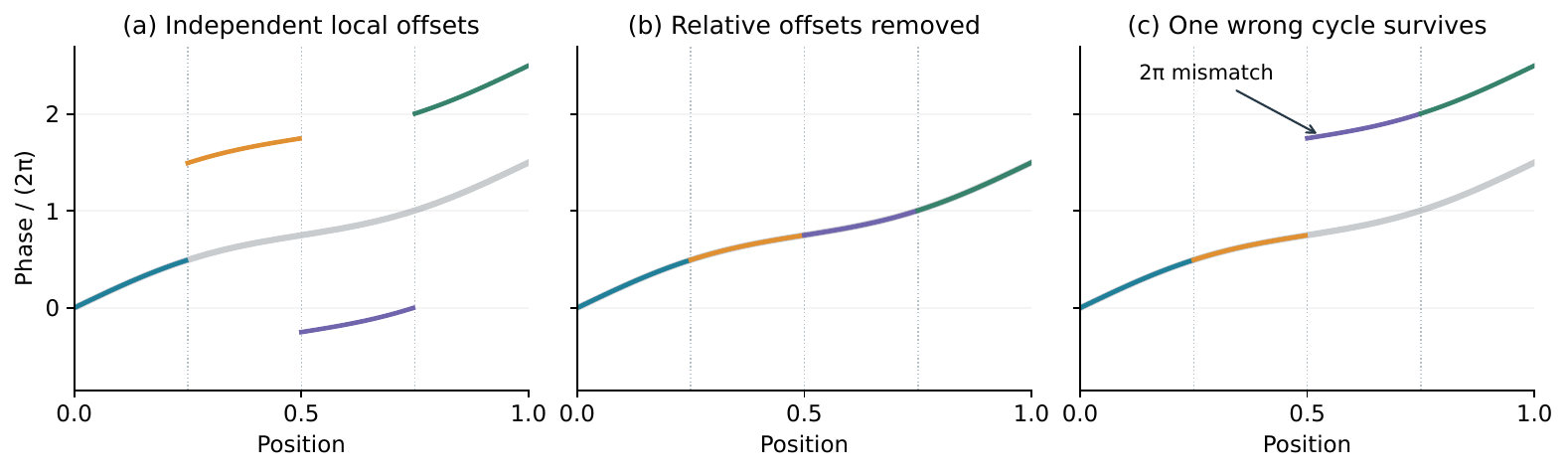}
\caption{\textbf{Three results with identical wrapped observations.} Each panel shows a one-dimensional cross-section through four units. Gray is the true phase, colors identify the units and dotted lines mark the unit boundaries. (a)~Each unit keeps its own offset: $0,+1,-1,+1$ cycles. (b)~The relative offsets are removed, and the result follows the true phase. (c)~One relative offset of one cycle remains between the two halves: the right half is wrong, although every unit wraps to the same observation.}
\label{fig:assembly}
\end{figure*}

\paragraph{Relative offsets and the common reference.}
A cycle shift common to all units moves the whole map but leaves the relations between units intact.
The overall level of the map, which such a shift changes, is called the common reference.
An evaluation should state how it was fixed before counting errors; otherwise, a common shift of one cycle counts as an error at every pixel.
The relative offsets between units are a different matter, as Figure~\ref{fig:assembly} shows.
A closed loop of units is a chain of neighboring units that returns to its start, such as four tiles around a corner.
The differences $c_j-c_i$ of one set of offsets always sum to zero around such a loop.
Differences measured on the seams need not, and when they do not, no single set of offsets agrees with all of them (Figure~\ref{fig:cycle}, Section~\ref{sec:across}).

\paragraph{Four distinct properties.}
The notions above show that four properties of a merged result are separate: a small value of the method's objective, congruence at every pixel, internal consistency, and accuracy against the true phase.
Internal consistency means that the offsets measured between units agree around every closed loop.
Each of these properties needs its own argument or measurement.

\section{Methods}
\label{sec:methods}
This section describes how the studies were found, selected, coded and checked.
The review follows the PRISMA 2020 reporting guideline for systematic reviews.\footnote{M.~J. Page et al., ``The PRISMA 2020 statement: an updated guideline for reporting systematic reviews,'' \emph{PLoS Medicine} 18(3), e1003583 (2021), \href{https://doi.org/10.1371/journal.pmed.1003583}{doi:10.1371/journal.pmed.1003583}.}
An exploratory search in September 2026 served to define the scope of the review and its queries.
The eligibility criteria below were fixed in a dated protocol on 1 October 2026, before the first author's screening, with later dated amendments.
On 29 October 2026, towards the end of the full-text reading, the same protocol recorded how the criteria had been applied to borderline cases (see Data availability).

\subsection{Eligibility criteria}
The review counts records, reports and studies.
A record is one entry returned by a search.
A report is one document, such as an article or a thesis.
A study is one piece of work, possibly described in several reports; its main report is its journal article when one exists.
A report is eligible when its full text meets three criteria.
First, it unwraps a phase defined on an image or a volume, in any modality; only one-dimensional unwrapping is excluded (I1).
Second, it partitions the domain into two or more units, fixed before the local solves that unwrap them, and it unwraps each unit at least partly on its own (I2).
Two kinds of units satisfy this criterion without a local solve of their own.
A region drawn so that it contains no jump needs no solve, because its phase is already continuous up to one integer, so it counts as unwrapped as it stands.
A partition imposed by the acquisition, such as subapertures, slices or scenes, also counts, because merging its units raises the same problem.
Third, it merges the unit solutions into one phase field, or states why its units must not be merged (I3).
A report that does not describe its merging stays included and is coded ``merging not reported'', because this gap is itself a finding.
Excluded are regions that grow during unwrapping, patches cut only to train a network, decompositions by frequency, time or radar baseline, resolution hierarchies without side-by-side units, reports that apply an existing unwrapper without describing a partitioned method, general reviews and retracted reports.
No restriction applies to publication date, language, document type or the presence of a DOI.

\subsection{Search}
OpenAlex, IEEE Xplore, Web of Science Core Collection and Scopus were searched on 1 October 2026.
The queries did not change afterwards.
Each query joined phase-unwrapping terms to partition and merging terms in titles and abstracts.
Appendix~\ref{app:search} gives the strings for each database.\phraseQueriesAgent{}
A citation search then started from \nSeeds database reports judged relevant on 1 October 2026, before the first author's screening, called seed reports; \nSeedsExcluded of them were later excluded at full text.
In OpenAlex, it collected the references of these seed reports and the documents that cite them, over one generation: the new candidates were not followed in turn.
Of the documents collected, a candidate was kept when its title or abstract contained the phase-unwrapping terms.
Finally, until 21 October 2026, identification by other methods added four kinds of candidates: records seen while testing the search strategy, documents from the first author's library and readings, a report listed in a thesis read in full, and the revised version of a report that could not be obtained.

\subsection{Selection}
\label{sec:selection}
Duplicates were removed on the normalized DOI, then, for records without a DOI, on the normalized title and on near-identical titles.
The same AI agent assisted the preliminary work of the selection.
It assisted the removal of duplicates and helped keep the registers up to date.
It also helped make a first, rapid screening of the titles and abstracts, on 1 and 2 October 2026.
The first author then screened every title and abstract, and recorded these decisions on 26 and 27 October 2026.
The second author independently screened a random sample of \nScreenSample database records, one fifth of them.
On this sample, the two authors disagreed on \nDisagreeScreening records, and settled these disagreements together.
The citation candidates were screened by the first author, without a second author, on the title for the \nCitNoAbstract without a published abstract.

A record retained by the first screening or by the first author was read in full.
On 29 October 2026, the criteria were amended to admit three-dimensional methods.
The screening was therefore repeated for these methods, which added \nKeptRescreen database records and \nCitKeptRescreen citation candidates.

For the retained records, full texts came from the institutional access of the Universit\'e de Montr\'eal, from open-access copies, from the first author's library or from interlibrary loan.
Some of these reports had already been read and annotated by the first author in earlier work, and these notes were reused.
The first author read every full text, focusing on the method sections, and decided its eligibility against criteria I1, I2 and I3.
The full texts were read progressively, from 1 to 30 October.
The eligibility decisions were recorded together on 29 October and revised until 5 November 2026.
The second author independently read a random sample of \nFullTextSample of these texts, one fifth of them, without seeing the first author's decisions.
The two authors made the same decision on \nFullTextAgree of these \nFullTextSample texts, and the first author's decision was kept for the other one.
The two included preprints by the authors went through the same procedure~\citep{moevus2026tilebased,moevus2026adaptive}.

\subsection{Data extraction, verification and synthesis}
Each included study has one row in the extraction table.
The row records the modality, the dimension, the partition family (fixed tiles, regions computed from the data, and so on), the units, the local solver, the merging rule and the merging order.
It also records three evidence items.
The first is a ground truth: an evaluation compares the result, by a number or on a figure, with a phase known independently of the method, from a simulation, a calibrated object or an independent measurement; another unwrapping algorithm does not count, nor does a reference that also served to calibrate the result.
The second is a comparison of the same local solver on the same image, run in one piece and on the units.
The third is the use of real data, acquired by an instrument rather than simulated.
Two further questions describe the merging: what it adjusts for each unit, and how these values are obtained (Section~\ref{sec:taxonomy}).
For \nAxesAmbiguous studies, two codes could be defended.
The counts use the code retained at extraction.
Both codes, and the counts obtained with the alternative ones, are available from the authors on request.
Finally, each study is coded as a new technique, a variant or combination, or the application of an existing technique.
An application was eligible when it described the partitioned processing well enough to assess I2 and I3, even if the technique was not new.

The statements that this review makes about the included studies were then checked against the full texts.
The same AI agent helped prepare the extraction table and make some of these verifications.
For \nPartialReading studies, only the chapters on the partition and the merging were read~\citep{wesolkowski2007stochastic,chen2013improving}.
Scanned PDFs and some PDFs in Chinese were read from page images.
Of the \nClaims statements, \nClaimsVerified were verified as written and \nClaimsReworded were reworded to match the text.

The synthesis is descriptive and counts studies, not reports.
Studies were grouped by mechanism into \nGroups groupings, which were proposed during extraction.
The groupings were then gathered into four classes, according to where the relation between units is decided.
A study belongs to the class of its dominant mechanism, and the studies that fit two classes are discussed in Section~\ref{sec:borderline}.
The counts of merging rules come from the two questions above, not from the groupings.
No effect measure was computed and no accuracy or runtime value was pooled, because the studies differ in their data, in what their scores are compared with, and in their measures.
No formal tool was used to assess the risk of bias or the certainty of the evidence; instead, the review records what each evaluation compares the results with, and whether it compares one piece with the units.
Reporting bias was not assessed formally either; the review counts instead the studies that do not report their merging.
No method was re-run, and the figures constructed for this article illustrate mechanisms only.

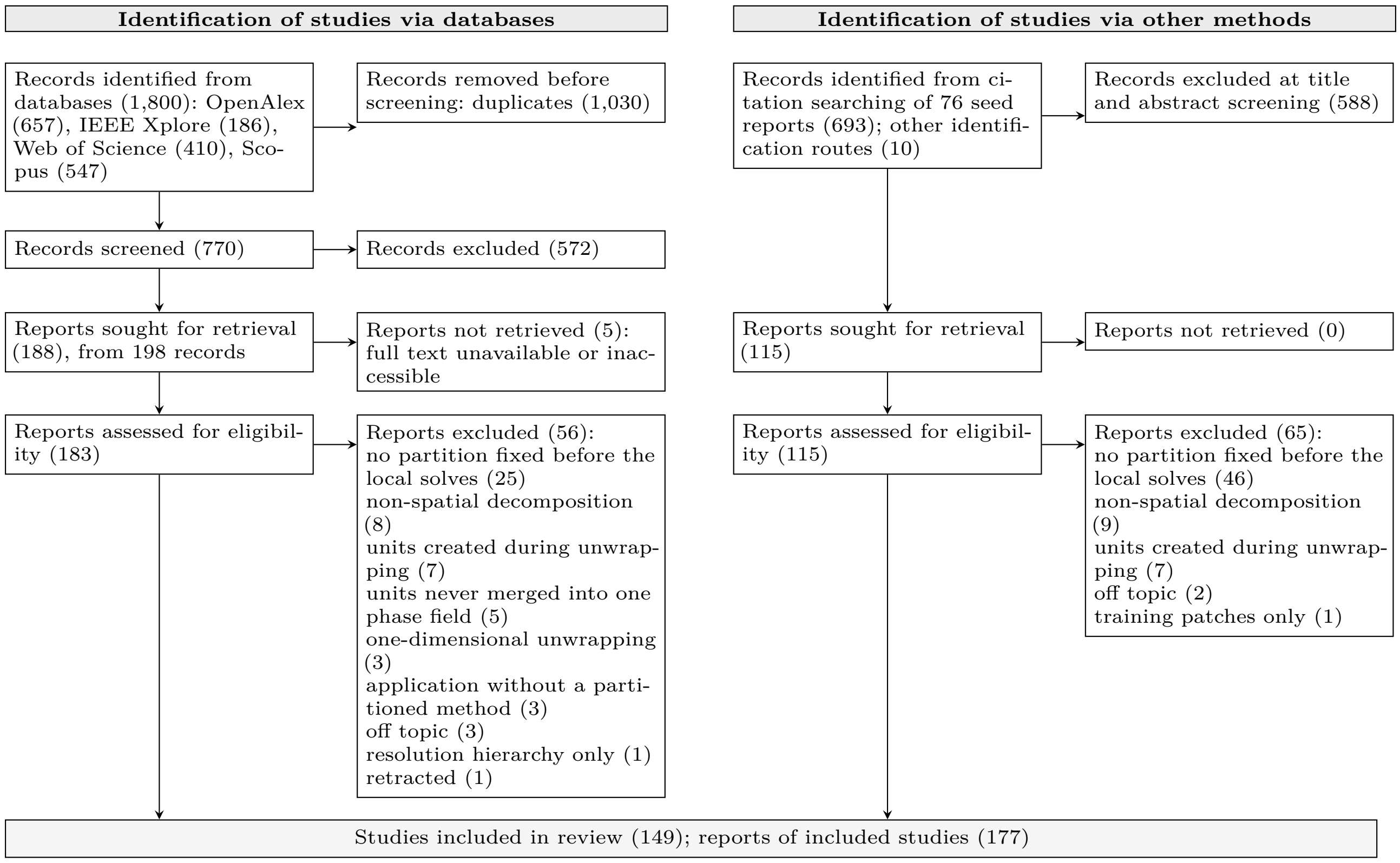
\begin{figure*}[tbp]\centering
\begin{tikzpicture}[font=\scriptsize, >=stealth,
  b/.style={draw, align=left, text width=3.55cm, inner sep=3pt, anchor=north west},
  h/.style={draw, fill=gray!15, align=center, text width=7.95cm, inner sep=2pt, font=\scriptsize\bfseries, anchor=north west}]
\node[h] (H1) at (0,0) {Identification of studies via databases};
\node[h] (H2) at (8.9,0) {Identification of studies via other methods};
\node[b] (a1) at (0,-0.7) {Records identified from databases (1,800): \mbox{OpenAlex} (657), IEEE Xplore (186), Web of Science (410), Scopus (547)};
\node[b] (a2) at (4.3,-0.7) {Records removed before screening: duplicates (1,030)};
\node[b] (a3) at (0,-2.75) {Records screened (770)};
\node[b] (a4) at (4.3,-2.75) {Records excluded (572)};
\node[b] (a5) at (0,-3.75) {Reports sought for retrieval (188), from 198 records};
\node[b] (a6) at (4.3,-3.75) {Reports not retrieved (5): full text unavailable or inaccessible};
\node[b] (a7) at (0,-5.0) {Reports assessed for eligibility (183)};
\node[b] (a8) at (4.3,-5.0) {Reports excluded (56):\\ no partition fixed before the local solves (25)\\ non-spatial decomposition (8)\\ units created during unwrapping (7)\\ units never merged into one phase field (5)\\ one-dimensional unwrapping (3)\\ application without a partitioned method (3)\\ off topic (3)\\ resolution hierarchy only (1)\\ retracted (1)};
\node[b] (c1) at (8.9,-0.7) {Records identified from citation searching of 76 seed reports (693); other identification routes (10)};
\node[b] (c2) at (13.2,-0.7) {Records excluded at title and abstract screening (588)};
\node[b] (c5) at (8.9,-3.75) {Reports sought for retrieval (115)};
\node[b] (c6) at (13.2,-3.75) {Reports not retrieved (0)};
\node[b] (c7) at (8.9,-5.0) {Reports assessed for eligibility (115)};
\node[b] (c8) at (13.2,-5.0) {Reports excluded (65):\\ no partition fixed before the local solves (46)\\ non-spatial decomposition (9)\\ units created during unwrapping (7)\\ off topic (2)\\ training patches only (1)};
\node[b, text width=16.55cm, align=center, fill=gray!8] (i1) at (0,-9.95) {Studies included in review (149); reports of included studies (177)};
\foreach \u/\v in {a1/a3, a3/a5, a5/a7, c1/c5, c5/c7} \draw[->] (\u.south) -- (\v.north);
\foreach \u/\v in {a1/a2, a3/a4, a5/a6, a7/a8, c1/c2, c5/c6, c7/c8} \draw[->] (\u.east) -- (\v.west |- \u.east);
\draw[->] (a7.south) -- (a7.south |- i1.north);
\draw[->] (c7.south) -- (c7.south |- i1.north);
\end{tikzpicture}
\caption{\textbf{PRISMA 2020 flow diagram.} Counts are computed from the eligibility registers of the review. One study can be described by several reports.}\label{fig:prisma}
\end{figure*}

\begin{table*}[tbp]
\centering\small
\caption{The corpus at a glance: the \nStudiesIncluded included studies, from the extraction verified against the full texts. Each study takes one value in each group of rows, except for the three evidence items, which are coded separately. The merging is described in Section~\ref{sec:taxonomy}.}
\label{tab:characteristics}
\begin{tabular}{@{}lr@{\hspace{3em}}lr@{}}
\toprule
\multicolumn{2}{@{}l}{\textit{Publication period}} & \multicolumn{2}{l@{}}{\textit{Partition family}}\\
\yearFirst--1999 & \nPeriodOne & Regions computed before unwrapping & \nFamRegions\\
2000--2009 & \nPeriodTwo & Fixed geometric tiles & \nFamFixed\\
2010--2019 & \nPeriodThree & Imposed by the acquisition & \nFamAcquisition\\
2020--\yearLast & \nPeriodFour & Adaptive tiles & \nFamAdaptive\\
\addlinespace
\multicolumn{2}{@{}l}{\textit{Modality}} & Two levels: coarse solution, then tiles & \nFamTwoLevel\\
Synthetic aperture radar or sonar & \nModSAR & Drawn by hand by the operator & \nFamManual\\
Optics and interferometry & \nModOptical & Other & \nFamOtherFamily\\
Profilometry and structured light & \nModProfilometry & \multicolumn{2}{l@{}}{\textit{Contribution}}\\
MRI & \nModMRI & New technique & \nApportNew\\
Holography and microscopy & \nModHolography & Variant or combination & \nApportVariant\\
Other & \nModOtherModality & Application of an existing technique & \nApportApplication\\
\addlinespace
\multicolumn{2}{@{}l}{\textit{Dimension}} & \multicolumn{2}{l@{}}{\textit{Evidence used in the evaluations}}\\
Two-dimensional & \nDimTwo & Ground truth & \nGroundTruth\\
Three-dimensional & \nDimThree & Same local solver in one piece and on the units & \nOneVsMany\\
$N$-dimensional & \nDimN & Real data & \nRealData\\
\bottomrule
\end{tabular}
\end{table*}

\section{Included studies}
\label{sec:included}
This section reports the result of the selection, then describes the included studies.

\subsection{Study selection}
\label{sec:flow}
Figure~\ref{fig:prisma} gives the flow of records and reports.
The database searches returned \nRecordsRaw records, and removing \nDuplicates duplicates left \nScreened.
Screening excluded \nExcludedScreening of them.
The \nRetainedRecords retained records describe \nRetainedWorks distinct documents, because some documents appear under more than one DOI.
The citation search screened \nCitScreened new candidates and retained \nCitRetained.

Full texts were then sought for \nSought reports: \nSoughtDB from the databases, \nSoughtCit from the citation search and \nSoughtOther from other methods.
Of these, \nNotRetrieved could not be obtained.
Four of them are conference abstracts or a software deposit with restricted files, for which no full article exists.
The fifth is a 1998 conference paper by Bo et al., outside the university's subscriptions; its revised 2001 journal version was read instead and counts as a report identified by other methods~\citep{bo2001weighted}.
All other texts were obtained.

Reading the full texts excluded \nExcludedFullText of the \nAssessed assessed reports (Table~\ref{tab:exclusions} in Appendix~\ref{app:search}); Appendix~\ref{app:excluded} lists each of them with its reason.
The most frequent reason, for \nExclSansPartition reports, was that no partition was fixed before the local solves: these reports solve the whole image in one piece, or use their blocks only as a mask, a filtering window, a grouping of residues for placing branch cuts or a schedule of the computation.
Another \nExclRegionsPendant reports form their regions during unwrapping, such as Baldi's region-growing method of 2003~\citep{baldi2003phase}.
Another \nExclObjetsSepares unwrap separate units as separate products, never meant to form one phase field.
In total, \nReportsIncluded reports were included, which describe \nStudiesIncluded studies.

\subsection{The corpus at a glance}
\label{sec:characteristics}
The selected studies can now be described as a whole.
The included studies were published from \yearFirst to \yearLast (Table~\ref{tab:characteristics}), most of them from 2010 onward.
Synthetic aperture radar or sonar is the most frequent modality (\nModSAR studies), followed by optics and interferometry (\nModOptical).
Only \nDimThree studies are three-dimensional and \nDimN $N$-dimensional.
Most studies are variants or combinations of existing techniques (\nApportVariant studies).
As for how the units are built, regions computed from the data before unwrapping come first (\nFamRegions studies), then fixed geometric tiles (\nFamFixed).

These counts, however, describe the included studies, not the whole field, and a frequent construction is not necessarily the most effective one.

\begin{figure*}[tbp]\centering
\definecolor{famA}{HTML}{0072B2}
\definecolor{famB}{HTML}{009E73}
\definecolor{famC}{HTML}{D55E00}
\definecolor{famD}{HTML}{CC79A7}
{\small\textbf{Partition-and-merge methods (149 studies): where is the relation between units decided?}}\par\medskip
\begin{tcbraster}[raster columns=4, raster equal height, raster column skip=2mm, raster left skip=0pt, raster right skip=0pt]
\begin{tcolorbox}[colback=famA!6, colframe=famA, boxrule=0.6pt, arc=2pt, left=3pt, right=3pt, top=2pt, bottom=2pt, title={\footnotesize\raggedright\hyphenpenalty=10000\exhyphenpenalty=10000 \textbf{I. Assembling locally unwrapped maps} \hfill 68\\\scriptsize\itshape decided after each unit is unwrapped}, fonttitle=\color{white}, coltitle=white]
{\scriptsize\renewcommand{\arraystretch}{1.05}\begin{tabularx}{\linewidth}{@{}>{\raggedright\arraybackslash\hangindent=0.6em}X@{\hspace{3pt}}r@{}}
\textbf{Merged one unit at a time} & \textbf{36}\\
\hspace*{0.5em}G1 Fixed grid, integer offset & 12\\
\hspace*{0.5em}G2 Fixed grid, real offset & 7\\
\hspace*{0.5em}G3 Adaptive or hierarchical partition & 9\\
\hspace*{0.5em}G4 Regions with a local solver & 8\\
\noalign{\vspace{2pt}}
\textbf{Merged jointly} & \textbf{16}\\
\hspace*{0.5em}G5 Offsets solved jointly & 16\\
\noalign{\vspace{2pt}}
\textbf{Units fixed by the acquisition or a network} & \textbf{16}\\
\hspace*{0.5em}G6 Acquisition: subapertures, slices, scenes & 11\\
\hspace*{0.5em}G7 Neural-network input size & 5\\
\noalign{\vspace{2pt}}
\end{tabularx}}
\end{tcolorbox}
\begin{tcolorbox}[colback=famB!6, colframe=famB, boxrule=0.6pt, arc=2pt, left=3pt, right=3pt, top=2pt, bottom=2pt, title={\footnotesize\raggedright\hyphenpenalty=10000\exhyphenpenalty=10000 \textbf{II. One integer per jump-free region} \hfill 37\\\scriptsize\itshape decided mostly without a local solve}, fonttitle=\color{white}, coltitle=white]
{\scriptsize\renewcommand{\arraystretch}{1.05}\begin{tabularx}{\linewidth}{@{}>{\raggedright\arraybackslash\hangindent=0.6em}X@{\hspace{3pt}}r@{}}
\textbf{Chosen region by region} & \textbf{33}\\
\hspace*{0.5em}G8 Jump-free regions, PRELUDE lineage & 19\\
\hspace*{0.5em}G9 Fringe regions, Gierloff type & 14\\
\noalign{\vspace{2pt}}
\textbf{Chosen by a graph energy} & \textbf{4}\\
\hspace*{0.5em}G10 Region labels by graph energy & 4\\
\noalign{\vspace{2pt}}
\end{tabularx}}
\end{tcolorbox}
\begin{tcolorbox}[colback=famC!6, colframe=famC, boxrule=0.6pt, arc=2pt, left=3pt, right=3pt, top=2pt, bottom=2pt, title={\footnotesize\raggedright\hyphenpenalty=10000\exhyphenpenalty=10000 \textbf{III. Consistency built into the solve} \hfill 23\\\scriptsize\itshape decided mostly during the solve}, fonttitle=\color{white}, coltitle=white]
{\scriptsize\renewcommand{\arraystretch}{1.05}\begin{tabularx}{\linewidth}{@{}>{\raggedright\arraybackslash\hangindent=0.6em}X@{\hspace{3pt}}r@{}}
\textbf{Constraints passed on} & \textbf{14}\\
\hspace*{0.5em}G11 Imposed values or coarse solution first & 5\\
\hspace*{0.5em}G12 Reliability regions, Dirichlet anchoring & 9\\
\noalign{\vspace{2pt}}
\textbf{Coordination during the solve} & \textbf{3}\\
\hspace*{0.5em}G13 Coordination during the solve & 3\\
\noalign{\vspace{2pt}}
\textbf{Cuts composed before integration} & \textbf{6}\\
\hspace*{0.5em}G14 Partition around residue clusters & 6\\
\noalign{\vspace{2pt}}
\end{tabularx}}
\end{tcolorbox}
\begin{tcolorbox}[colback=famD!6, colframe=famD, boxrule=0.6pt, arc=2pt, left=3pt, right=3pt, top=2pt, bottom=2pt, title={\footnotesize\raggedright\hyphenpenalty=10000\exhyphenpenalty=10000 \textbf{IV. Units related by a reference, or kept apart} \hfill 9\\\scriptsize\itshape decided outside the seams}, fonttitle=\color{white}, coltitle=white]
{\scriptsize\renewcommand{\arraystretch}{1.05}\begin{tabularx}{\linewidth}{@{}>{\raggedright\arraybackslash\hangindent=0.6em}X@{\hspace{3pt}}r@{}}
\textbf{Merging against an external reference} & \textbf{8}\\
\hspace*{0.5em}G15 External absolute reference & 8\\
\noalign{\vspace{2pt}}
\textbf{Merging refused on purpose} & \textbf{1}\\
\hspace*{0.5em}G16 Merging refused (physical discontinuity) & 1\\
\noalign{\vspace{2pt}}
\end{tabularx}}
\end{tcolorbox}
\end{tcbraster}
\par\smallskip\begin{tcolorbox}[colback=gray!6, colframe=gray!60, boxrule=0.6pt, arc=2pt, left=3pt, right=3pt, top=2pt, bottom=2pt, title={\footnotesize\textbf{G17. Partition described, merging not described} \hfill 12}, fonttitle=\color{white}, coltitle=white]
{\scriptsize\itshape Shown apart from the four classes: these studies describe their partition but not their merging.\par}
\end{tcolorbox}
\caption{\textbf{Taxonomy of the \nGroups groupings.} The four classes differ in where the relation between units is decided. Numbers count studies. G1 to G17 are listed in Table~\ref{tab:groupings}. A few studies do not fit their class neatly (Section~\ref{sec:borderline}). G17 holds the studies whose merging is not described well enough to be assigned; other reporting gaps occur within the four classes.}\label{fig:hierarchy}
\end{figure*}

\section{A taxonomy of partition-and-merge methods}
\label{sec:taxonomy}
This section presents the main result of the review.
The studies form \nGroups groupings (Table~\ref{tab:groupings} in Appendix~\ref{app:included}).
All groupings except G17 fall into four classes, according to where the relation between units is decided (Figure~\ref{fig:hierarchy}).
G17 stays apart, because its \nGroupNR{} studies do not describe their merging.
In class~I, the relation is decided after the units have been unwrapped: each unit becomes a phase map of its own, and the maps are then shifted into place.
In class~II, the regions are drawn so that, by assumption, they contain no jump.
The merging then chooses one integer per region, mostly without a local solve.
In class~III, the relation is mostly decided during the solve.
Three mechanisms keep the units consistent: values passed on between solves, cooperation between units, and branch cuts combined before one integration.
In class~IV, the relation does not come from comparing units across their seams: an external reference places the units, or the merging is refused across a physical discontinuity.

Three remarks help to read the figure and the rest of this section.
First, a study belongs to the class of its dominant mechanism, and a few studies do not fit neatly (Section~\ref{sec:borderline}).
Second, inside class~I, the figure sorts most groupings by their merging, one unit at a time or jointly.
Groupings G6 and G7 are defined differently: they gather the studies whose units are fixed by the acquisition or by a network, whatever their merging.
Third, ``first'', ``earliest'' and ``only'' refer to the included studies.
An earliest study is the one whose main report is the oldest.

Two questions, coded for every study, describe the merging across the classes.
The first question is what the merging adjusts for each unit.
This review calls this quantity the merging variable.
The most frequent answer is one integer multiple of $2\pi$ (\nVarInteger studies).
One real-valued offset is adjusted in \nVarReal studies and another variable in \nVarOther, while nothing is adjusted after the local solves in \nVarNone.
The second question is how the values of this variable are obtained.
In \nOrgOneAtATime studies, they are set one unit at a time, along an order, a tree or a hierarchy of pairs, and in \nOrgJoint they are solved all together.
The other studies relate their units in another way, or not at all: \nOrgDuring keep the units consistent during the solve, \nOrgSingleIntegration prepare all units before one integration, \nOrgReference place each unit against an external reference, and \nOrgNotApplicable never merges its units.

Not every study describes its merging.
Grouping G17 holds the \nGroupNR{} studies that describe their partition but not their merging, or defer it to another source.
Three of them still give a merging order or state that an offset is applied, without saying how it is estimated~\citep{huang2007phase,madhu2007new,gao2017twodimensional}.
Some studies placed in other groupings also leave part of the merging out.
In all, \nVarNotReported studies therefore do not say what their merging adjusts, and \nOrgNotReported do not say how the values are obtained.
An unreported merging rule is a reporting gap, not evidence that merging is absent.

\begin{figure*}[tbp]
\centering\includegraphics[width=\linewidth]{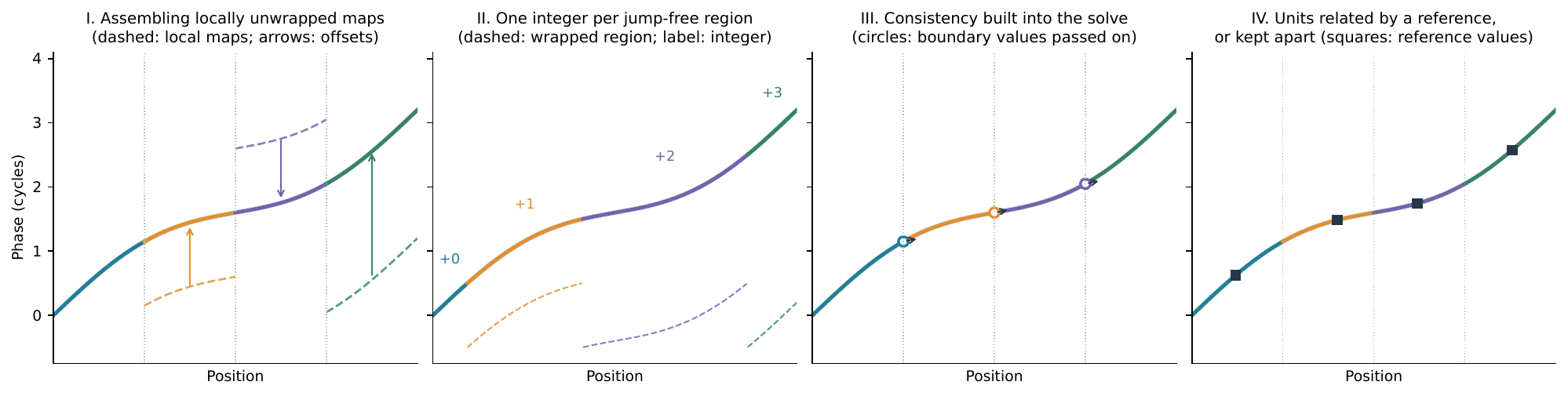}
\caption{\textbf{A typical mechanism of each class, on one constructed profile.} The solid curve is the true phase, in cycles; colors identify units. (I)~Each tile is unwrapped on its own, with an arbitrary offset (dashed), then shifted into place (arrows). (II)~Regions are cut where the wrapped phase jumps; each region keeps its wrapped values (dashed) and receives one integer. (III)~Each unit is solved with the boundary value already fixed by its neighbor (circles). (IV)~Each unit is placed by an external measurement (squares). The sketches do not cover every study of a class.}
\label{fig:classes}
\end{figure*}

Figure~\ref{fig:classes} sketches a typical mechanism of each class on one constructed profile.
Sections~\ref{sec:classI} to~\ref{sec:classIV} describe the four classes, and Section~\ref{sec:across} gathers the issues raised by merging across seams.

\subsection{Class I: assembling locally unwrapped maps}
\label{sec:classI}
\begin{figure*}[tbp]
\centering\includegraphics[width=\linewidth]{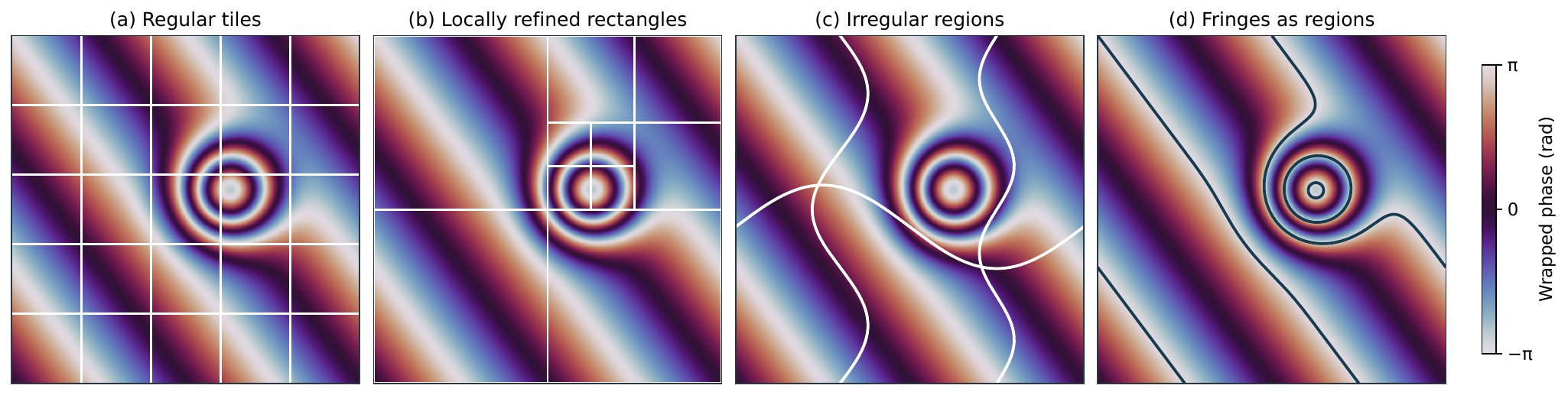}
\caption{\textbf{One observation, several possible units.} Illustrative partitions over the same synthetic wrapped phase: (a)~a regular grid, (b)~a local refinement into smaller rectangles and (c)~irregular regions. (d)~The fringes used as regions, as in class~II. Panels (a) to (c) show units of class~I, and white lines mark their boundaries. In (d), dark lines mark the jumps of the wrapped phase, which bound the fringes. Colors represent the wrapped phase.}
\label{fig:partition}
\end{figure*}

In class~I, the units are fixed tiles, adaptive or hierarchical tiles, regions unwrapped by a local solver, or units fixed by the acquisition or a network.
Figure~\ref{fig:partition} shows such constructions on one observation.
This section first describes these four kinds of units, then three ways of merging them: one unit at a time, by averaging overlapping estimates, and all units jointly.
The paragraphs on units also cite studies of other classes that build the same kind of units.

\paragraph{Fixed tiles.}
A fixed grid makes the size, the overlap and the boundaries of the units explicit, but it does not fix what the units return.
What each unit returns is set by the local solver, and the studies differ on this point.
Strand et al. unwrap each block by a constant phase rotation, assuming at most one jump per block~\citep{strand1999two}.
This rotation adds a constant to the wrapped phase, wraps the sum again and subtracts the constant.
Instead of a rotation, Antonopoulos et al. fit a polynomial to each tile~\citep{antonopoulos2015tilebased}.
Friedlander and Francos also fit a polynomial to each block, and they cut the field into 16 blocks in their example~\citep{friedlander1996model}.
Schwarz uses a simpler model of the same kind: he reduces each tile of a half-overlapping grid to an affine plane~\citep{schwarz1999hybrid}.

The overlap between tiles can serve two purposes.
First, it can supply the observations used to relate neighboring tiles.
Hao et al. choose the multiple of $2\pi$ of each tile on its 50-pixel overlaps with the tiles already placed~\citep{hao2017phase}.
Towers et al. use their 4-pixel overlap differently: they compare the edge profiles of neighboring tiles to score how well the tiles fit.
This score does not set the offset of a tile: it helps to decide the order in which the tiles are merged~\citep{towers1991automatic}.
Second, the overlap can serve only to blend outputs: Song et al. (2026) average the outputs of a neural network with a smooth window over a 128-pixel overlap~\citep{song2026insar}.

The tiles that make a computation possible need not be the units that are merged.
Chen and Zebker tile a large interferogram to run their network-flow solver SNAPHU within its computing limits.
They then segment each tile into reliable regions and merge these regions, not the tiles~\citep{chen2002phase}.
Two parts of one tile can therefore move relative to each other.
Carballo and Fieguth also merge regions.
They unwrap overlapping groups of $2\times2$ fixed rectangles, so that each pixel is unwrapped four times.
They keep a region only if its four unwrappings differ by a constant~\citep{carballo2002hierarchical}.

A fixed grid also makes assumptions about the phase.
Costantini advises blocks large enough that wrong first estimates of the differences between neighboring pixels do not split them into disconnected regions~\citep{costantini1998novel}.
In the simplest choice that he gives, these first estimates are the wrapped differences.
Gao et al. (2017) advise a block size that does not fragment discontinuity boundaries~\citep{gao2017twodimensional}.
When such an assumption fails, Hao et al. repair the grid: if a discontinuity zone crosses several tiles, they merge these tiles into a larger tile and unwrap it again~\citep{hao2017phase}.

\paragraph{Adaptive and hierarchical tiles.}
An adaptive partition fits the size or the shape of its units to the content of the image.
Baldi splits an area into four quarters as long as one of its pixels has a neighbor with a phase difference above $\pi$~\citep{baldi2001twodimensional}.
These nested splits into four form a quadtree, the first adaptive one among the included studies.
Each group of four sub-areas is then merged with three integer offsets.
Other adaptive partitions follow other features of the image.
Before Baldi, Wang et al. (1999) let the blocks follow masks, edges and phase jumps~\citep{wang1999phaseunwrapping}.
Later, Huang et al. (2006) move the sides of overlapping segments according to the geometry of the valid pixels~\citep{huang2006phase}.
Sai Prasad et al. follow the object rather than the phase: they choose the tiles from the geometry of the specimen, in a method they describe as interactive~\citep{saiprasad2004towards}.
Moevus and Mignotte return to the phase and refine quadtree and k-d tree partitions where residues lie~\citep{moevus2026adaptive}.
A k-d tree splits a rectangle in two with one horizontal or vertical cut.
Section~\ref{sec:evidence} reports their comparison with a regular grid.

A hierarchy fixed in advance is not an adaptation, because its tree ignores the phase.
The earliest such hierarchy is that of Arevalillo Herr\'aez et al. (1996).
They cut the image recursively into four, down to blocks of $2\times2$ pixels, unwrap each part on its own and merge the parts level by level~\citep{arevalilloherraez1996robustsimple}.
Their conference report of the same year builds the hierarchy from the bottom instead: it groups small tiles, for example of $4\times4$ pixels, into larger and larger groups~\citep{arevalilloherraez1996robust}.
Hardie et al. also fix in advance a binary tree of blocks, from pixel pairs up to the two image halves, and merge adjacent blocks bottom-up~\citep{hardie1998robust}.
Chen et al. (2025) likewise nest regular blocks of $2^j\times2^j$ pixels, where $j$ is the level of the grid, up to the whole image~\citep{chen2025phase}.
Wu, Feng and Liu also merge the blocks of a fixed grid bottom-up, in groups of $2\times2$ and then in pairs~\citep{wu2007phase}.

\paragraph{Regions unwrapped by a local solver.}
Some partitions draw regions that still contain jumps, so each region is unwrapped by a local solver and merged like a tile.
Lu and Su draw such regions from the data: they threshold a quality map of phase-derivative variance into several quality regions, three in their example~\citep{lu2014fast}.
This quality map measures, in a window around each pixel, how much the derivatives of the wrapped phase spread around their mean.
Li et al. draw them from the physics instead.
They cut an interferogram along the trace of a surface rupture and unwrap each part separately.
They then merge the two parts on their overlap, where more than 99.86\,\% of the pixels already agree~\citep{li2015coseismic}.

\paragraph{Units fixed by the acquisition or a network.}
In other studies, the acquisition imposes the partition.
In subaperture testing of optics, each measurement of a part of the surface gives one unit.
In simulation, Lin et al. unwrap each circular subaperture by fitting 36 Zernike polynomials, a standard family of smooth shapes defined on a disk.
They then stitch the overlapping subapertures, which are arranged in rings~\citep{lin2011subaperture}.
On a real mirror, Zhao et al. (2014) move the tested part for each subaperture and require an overlap of at least a quarter of the mirror area~\citep{zhao2014study}.
The acquisition also sets the units in other imaging techniques.
In scanning broadband-light interferometry, each scan position supplies one unit~\citep{zhou2017topography,zhao2022high}.
In MRI, Liu and Drangova unwrap each slice of a volume separately, then stack the slices with one integer per slice.
A second step unwraps planes that contain the slice direction~\citep{liu2012interventionbased}.
The shape of the instrument can also set the units: Nie et al. draw eight fixed regions on the simulated pupil of a mirror made of 18 hexagonal segments, following the shape of the pupil rather than the data~\citep{nie2018study}.
Finally, Ramesh adapts the acquisition to the fringes: the zones of high fringe density are acquired at a higher optical magnification~\citep{ramesh2000phase}.

These studies are grouped by how their units are built, whatever their merging, so their merging is described here.
It often uses real-valued geometric parameters rather than cycles.
Zhao et al. (2014) remove a constant (piston), two slopes (tilts) and a quadratic term (defocus), estimated against a reference subaperture~\citep{zhao2014study}.
By contrast, Xie et al. (2021) estimate only a lateral translation between unwrapped maps~\citep{xie2021phase}.
In airborne radar mapping, Gao et al. (2022) estimate three parameters for each radar scene: the baseline length, the baseline angle and an absolute phase offset.
The baseline is the vector from one antenna of the radar to the other.
They estimate these parameters by a joint least-squares fit on ground control points and on points shared by overlapping scenes~\citep{gao2022experimental}.

A neural network can also impose the units, through its fixed input size.
Zhao et al. (2021), who describe the earliest such tiling, cut the wrapped phase into $256\times256$ subregions with more than 20\,\% overlap~\citep{zhao2021phase}.
Their merging works on the phase values themselves: it registers neighboring tiles and merges their overlap by a weighted average, as described below.
Liu et al. (2024), by contrast, register neighboring tiles on the parity of the number of cycles, the only included study to do so~\citep{liu2024highaccuracy}.
Each tile returns a binary map of this parity.
An exclusive-or, the operation that tells whether two binary values differ, is applied to two adjacent maps.
It decides whether the values 0 and 1 of one map must be inverted before the two maps are merged.

\paragraph{Merging one unit at a time.}
Merging one unit at a time raises two questions: whether the merged field stays congruent with the wrapped data, and in which order the units are fixed.
Congruence is treated first, and the order afterwards.
Whether the merged field stays congruent depends on the local output and on the values allowed for each offset (Section~\ref{sec:background}).
Strand et al. keep congruence throughout, because their rotation changes each block only by multiples of $2\pi$, and each block receives the rounded mean of the differences across the seam~\citep{strand1999two}.
Several other studies do not have this property.
They start from least-squares local solves, which are generally not congruent (Section~\ref{sec:background}), and add real-valued offsets.
For example, Chen, Wang and Zhang add the mean phase difference over the overlap, without rounding~\citep{chen2025stitching}.
Wang et al. (2013) go from integers to reals: they replace the rounded multiple of $2\pi$ of an earlier method with a real mean offset~\citep{wang2013improvement}.
Moevus and Mignotte also use one real-valued offset per tile~\citep{moevus2026tilebased}.
In these descriptions, no step restores congruence with the wrapped data at every pixel.

A local model can restore congruence by construction.
Antonopoulos et al. fit a polynomial $f$ to the wrapped differences of a tile and return, schematically, $u=f+\W(\psi-f)$~\citep{antonopoulos2015tilebased}.
The wrapped residual $\W(\psi-f)$ makes $u$ congruent at every pixel, and integer offsets preserve this property.
Friedlander and Francos apply the same correction after merging: each pixel takes the multiple of $2\pi$ that brings it closest to the block model~\citep{friedlander1996model}.
A low fitting residual and a correct multiple of $2\pi$ at each pixel remain two separate claims (Section~\ref{sec:background}).

The order of merging decides which seams set each offset.
A sequential order fixes each unit once, against the area already merged, as in the raster scans of Strand et al., Huang and He, and Hao et al.~\citep{strand1999two,huang2002phase,hao2017phase}.
An error fixed early is then passed on to every unit merged through it.
Chang et al. state that their merging depends on the path when the phase map contains inconsistencies~\citep{chang1998tapestry}, and Carballo and Fieguth reject sequential merging because an incorrectly unwrapped block affects all the rest~\citep{carballo2002hierarchical}.

A reliability order pushes doubtful units to the end.
Towers et al. give the earliest such merging~\citep{towers1991automatic}.
They treat each tile as a node of a graph, with an edge between neighboring tiles.
They merge along the minimum spanning tree of this graph, that is, the set of edges of smallest total weight that connects all tiles without a loop.
The edge weight combines the confidence of the overlap match, the fringe density and the number of low-modulation pixels, so that defective tiles end at the tips of the branches.
Later studies also order the merging by reliability.
Schwarz grows a region of tiles by adding, at each step, the tile that matches it best~\citep{schwarz1999hybrid}.
Antonopoulos et al. apply to tiles the reliability ordering of SRNCP (sorting by reliability following a noncontinuous path), a pixel-level unwrapper~\citep{arevalilloherraez2002fast,antonopoulos2015tilebased}.
Block or seam quality also guides the order in several other studies~\citep{lo2002surface,huang2007phase,dalmaucedeno2008fast,huang2010fast,zhong2019hierarchical,chen2025stitching}.
Moevus and Mignotte also merge along a spanning tree, but their weights grow as the measurements on a seam agree.
They therefore propagate offsets along a maximum-weight spanning tree~\citep{moevus2026tilebased,moevus2026adaptive}.

A tree can also fix a hierarchy of merges: small groups of units first, then groups of groups.
Arevalillo Herr\'aez et al. (1996) bring the four zones of a group together, one pair of neighboring zones at a time.
They add to the second zone of each pair the multiple of $2\pi$ on which the most seam points agree, and mask both zones when this agreement is too rare~\citep{arevalilloherraez1996robustsimple}.
Wu, Feng and Liu also merge groups of four, but with real values: they give each block of a $2\times2$ group a real constant, estimated by weighted least squares on the border pixels of the group~\citep{wu2007phase}.
A correction can also follow the merging: Huang et al. (2006) correct thin strips against a polynomial fitted to the whole map~\citep{huang2006phase}.

\paragraph{Overlapping estimates averaged.}
Overlapping estimates can be averaged instead of shifted.
Moevus and Mignotte repeat their tiled unwrapping for 16 positions of the grid and the 8 rotations and reflections of the image, which gives 128 unwrapped results in all~\citep{moevus2026tilebased}.
They average these results with a weight that decreases with a residual score.
This score compares the Laplacian of each result with that of the observation.
The Laplacian combines, at each pixel, the differences with its four neighbors.
Here both Laplacians are computed from neighbor differences from which full-cycle jumps are removed.
It does not guarantee the correct number of cycles.
Zhao et al. (2021) also average, but only on the overlaps: they register network tiles by an iterative-closest-point method on $(x,y,\text{phase})$ point clouds, and then merge the overlap by a weighted average~\citep{zhao2021phase}.

\paragraph{Merging all units jointly.}
A joint solution puts all seam relations into one system, so that conflicting relations become visible.
Hung and Yamada are the earliest to specify such a solve~\citep{hung1998phase}.
Before them, Gierloff already infers the final offsets from the comparisons between all regions before assigning them, but does not specify the algorithm~\citep{gierloff1987phase}.
In the method of Hung and Yamada, each region boundary gives one equation on the difference between two region integers.
When ambiguous points make the system inconsistent, it is solved by least squares, weighted by the number of pixel pairs on each boundary, and the solution is rounded to integers.

A network flow can also solve all integer offsets jointly.
Among the included studies, two studies of 2002, published in the same journal issue, are the first to do so: Carballo and Fieguth, and Chen and Zebker~\citep{carballo2002hierarchical,chen2002phase}.
Chen and Zebker credit the thesis of Carballo (2000) for the principle of a second flow problem between reliable regions.
They pose this second network-flow problem between regions instead of neighboring pixels.
An arc of a flow network is a link between two nodes, along which flow can pass.
The arcs of the second problem follow region boundaries.
Each of these arcs costs the sum of the costs of the pixel-level arcs along it.
The nonlinear solver of SNAPHU solves this second problem.
Gao et al. (2005) also solve a second network flow on a Delaunay triangulation of one point per tile~\citep{gao2005phase}.
A Delaunay triangulation is a standard way to link scattered points into triangles.
Wu et al. (2024) base their block merging on that of Chen and Zebker, with changes that they do not detail~\citep{wu2024parallel}.
Zhong et al. (2010) use another joint criterion, which is not a flow: they choose one integer per block by minimizing the discontinuities along block borders, then finish with a minimum-discontinuity step over the whole image~\citep{zhong2010fast}.

Integer programs pose the same joint problem in other forms.
Kelany et al. pose the choice of one integer offset per sub-image as a quadratic unconstrained binary optimization (QUBO), which minimizes a quadratic function of binary variables~\citep{kelany2022quantum}.
Haghighi and Dimopoulos pose the problem differently: they solve one equality constraint per pair of neighboring sub-images with a constrained quadratic model (CQM)~\citep{haghighi2024enhanced}.
Jiang et al. reach integers by another route, without an integer program.
They predict the number of cycles of each pixel with a network on overlapping tiles, then solve all tile offsets jointly by real weighted least squares and round the solution~\citep{jiang2026wrapcountbased}.

Some joint solutions drop the integer constraint.
Yang and He solve a real-valued scale and offset per tile on the overlap graph~\citep{yang2026continuous}.
Gourdain and Bachmann also keep real constants: they split a radial-basis-function network into sub-networks and fit one real constant per sub-network on shared ghost neurons, boundary neurons duplicated in exactly two sub-networks~\citep{gourdain2024neural}.
Lin et al. combine a joint and a sequential step: they merge the subapertures of each ring by one least-squares fit of piston and tilts, and then merge the rings one at a time~\citep{lin2011subaperture}.

\subsection{Class II: one integer per jump-free region}
\label{sec:classII}
In class~II, the regions are drawn before unwrapping so that, by assumption, none contains a jump.
Most methods therefore need no local solver.
The absence of jumps is an assumption, not a guarantee: in the PRELUDE method described below, a region can join two areas whose true phases differ by more than $2\pi$~\citep{jenkinson2003fast}.
In many class~II methods, these regions are the fringes themselves, detected before any unwrapping (Figure~\ref{fig:partition}(d)).

\paragraph{Drawing the regions.}
Regions differ mainly by the signal that draws their boundaries.
In the oldest included study, Gierloff forms all jump-free regions on the wrapped phase before deciding any offset~\citep{gierloff1987phase}.
Huang and Lai also start from the wrapped phase: they identify their jump-free sub-maps by flood fill, a technique that spreads one label from a pixel to all the connected pixels of its region~\citep{huang2002innovative}.
In their recommended mode, a few iterations of a local $\pm2\pi$ rule come before this flood fill (Section~\ref{sec:borderline}).
Charette and Hunter follow Gierloff but draw the regions from the quality of local plane fits~\citep{charette1996robust}.
Stetson et al. use a binary signal, one label per pixel: whether its phase lies within $\pm\pi/2$~\citep{stetson1997noiseimmune}.
Arevalillo Herr\'aez et al. (2002) also use intervals of the wrapped phase: they group connected pixels whose wrapped phase falls in one interval~\citep{arevalilloherraez2002robust}.
PRELUDE (Phase Region Expanding Labeler for Unwrapping Discrete Estimates) builds its regions on the same principle and extends them to volumes~\citep{jenkinson2003fast}.
SEGUE (Speedy rEgion-Growing algorithm for Unwrapping Estimated phase) also builds three-dimensional regions, from six phase intervals~\citep{karsa2019segue}.
CLOSE (pixel clustering and local surface fitting), an MRI method, draws blocks where the phase is smooth, then jump-free sub-blocks within each block~\citep{cheng2017close}.
A second measurement can also draw the boundaries: Wang et al. (2021) segment the difference between the phases measured at two modulation frequencies~\citep{wang2021phase}.

\paragraph{Choosing the integers region by region.}
Region methods often merge greedily on the graph that links adjacent regions: at each step, they make the merge that looks best and never revise it.
Charette and Hunter shift a region by $\pm2\pi$ when the mean difference across its boundary exceeds a threshold, and then set the offsets from the largest region outward~\citep{charette1996robust}.
The other methods differ in the order of the merges.
PRELUDE first merges the pair whose one-cycle error would cost the most and stops after $N-1$ merges for $N$ regions~\citep{jenkinson2003fast}.
SEGUE favors large boundaries instead: it lets the region with the largest boundary absorb all its accepted neighbors at once~\citep{karsa2019segue}.
Arevalillo Herr\'aez et al. (2002) and Xiong et al. also favor long boundaries, and resolve the longest boundary first~\citep{arevalilloherraez2002robust,xiong2008phase}.
Stetson et al. use boundary length the other way around: they start with the smallest regions, and merge each into the neighbor with which it shares the longest boundary~\citep{stetson1997noiseimmune}.
CLOSE relies on a local model instead: it uses a local quadratic fit to choose the integer shared by each sub-block, then by each block~\citep{cheng2017close}.
Whatever the order, greedy merging carries no guarantee of a global integer optimum.

\paragraph{Choosing the integers by a graph energy.}
Integer labels can also be chosen by minimizing an energy on the region graph.
This energy is a cost that sums one term per region and one term per pair of adjacent regions.
Xie et al. (2010) choose labels in $\{-1,0,+1\}$ for all patches at once, by graph-cut optimization~\citep{xie2010patchbased}.
Graph-cut optimization minimizes such an energy by cutting a graph whose edge weights encode the energy.
Yang et al. also assign one fringe order per region, with a per-region term from stereo matching of a projected speckle pattern~\citep{yang2025phaseconnectedregionbased}.
Wang et al. (2021) call their model a Markov random field, but they choose the integers node by node~\citep{wang2021phase}.
Each node takes the pair of integers that scores best for its own terms, so the choice is local rather than a minimum of the whole energy.
The order of the nodes is not given.

\subsection{Class III: consistency built into the solve}
\label{sec:classIII}
\begin{figure*}[tbp]
\centering\includegraphics[width=\linewidth]{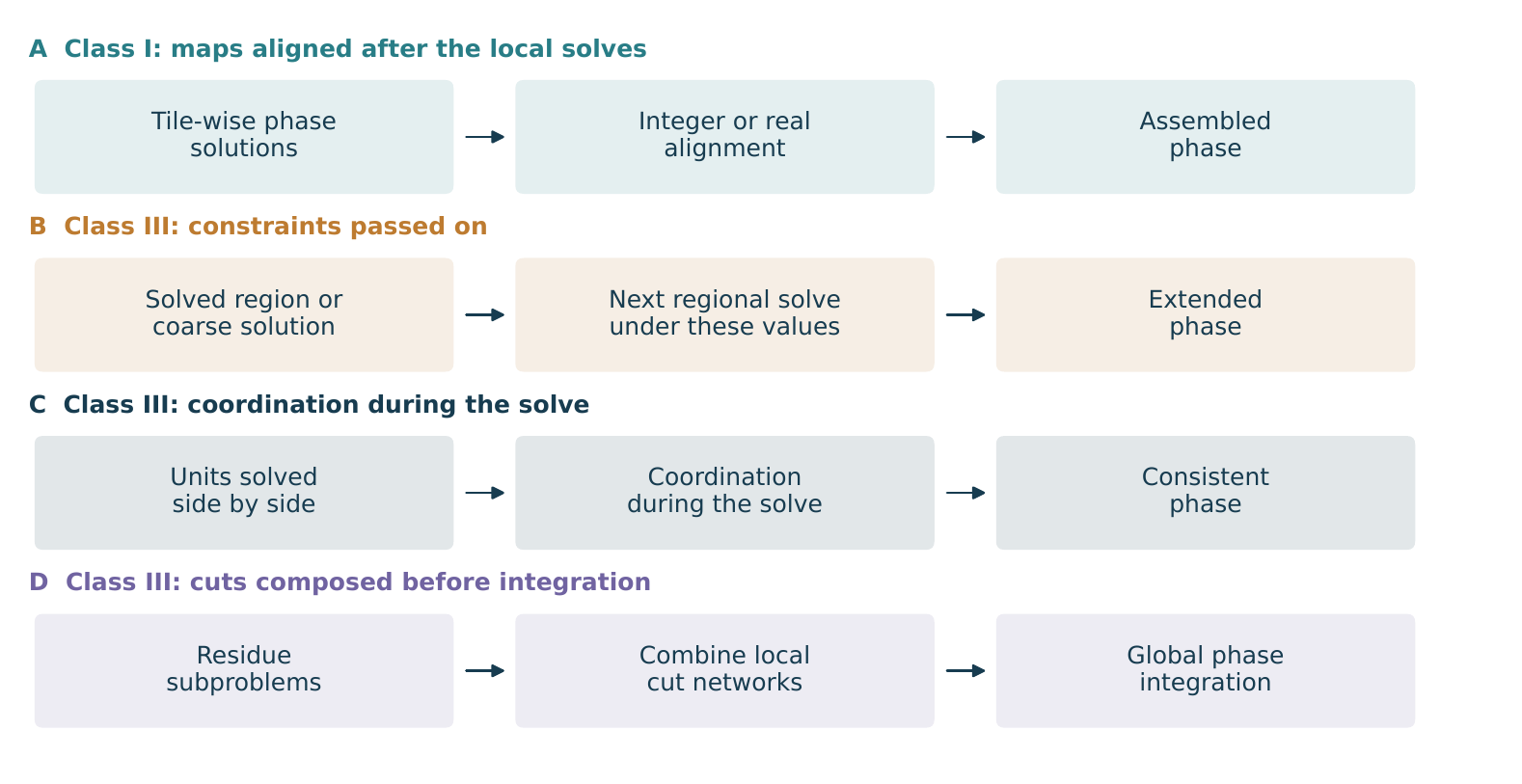}
\caption{\textbf{The three mechanisms of class~III, compared with class~I.} (A)~Local phase maps aligned after the local solves (class~I). (B)~Each region solved under values already fixed, by the regions solved before or by a coarse solution. (C)~Units solved side by side and coordinated during the solve, typically by exchanging shared values. (D)~Local cut decisions combined before one integration. Boxes represent steps, not equally costly operations.}
\label{fig:architectures}
\end{figure*}

In class~III, the units mostly do not need to be reconciled afterwards, because consistency is built into the way they are solved.
Figure~\ref{fig:architectures} compares the three mechanisms of class~III (rows B to D) with the alignment of class~I (row A).

\paragraph{Constraints passed on.}
In the first mechanism (row~B of Figure~\ref{fig:architectures}), a unit is solved with some of its values already fixed, by the regions solved before or by a first solution.
Most of these methods therefore estimate no offset.
Fornaro and Sansosti partition the image into regions and give each region a priority value, which sets the order in which the regions are solved~\citep{fornaro1999twodimensional}.
They draw these priorities from the coherence, which measures how reliable the phase of each pixel is, and from layover, a distortion of radar images on steep slopes.
The most coherent region is solved first, and the layover areas receive the lowest priority.
Each new region is solved by least squares.
On its border with the regions already unwrapped, its values are fixed to the values of these regions (Dirichlet conditions).
No step restores congruence with the wrapped phase.
The same scheme reappears with geometric blocks and in finite-element solvers~\citep{fornaro2001generation,wu2004insar,zhang2009new}.

The fixed values can also be integer flows, the whole numbers of cycles that a network-flow method adds to the wrapped differences (Section~\ref{sec:background}).
Costantini gives the first merging of this kind: overlapping blocks are solved one after another by minimum-cost flow, and the flows on each overlap are fixed to the solution of the blocks already processed~\citep{costantini1998novel}.
He calls this block solution suboptimal.
Xu et al. fix values rather than flows: they isolate low-quality tiles by thresholds and unwrap only these tiles~\citep{xu2015phase}.
The border of each tile touches areas judged unambiguous, so the unwrapping leaves the phase on this border unchanged.

A first solution can also be computed before the units are drawn, on part of the pixels or in an approximate form.
Pepe et al. first unwrap a sparse set of highly coherent pixels.
They keep the pixels of this set that they judge successfully unwrapped, and join them by a Delaunay triangulation, a mesh of triangles whose corners are these pixels.
The edges of this mesh then cut all coherent pixels into tiles, and the borders of these tiles carry this first solution~\citep{pepe2011new}.
In their experiment, this first network holds about 50{,}000 points, out of about 700{,}000 coherent pixels.
Zhang et al. (2011) compute an approximate $L^1$ solution of the whole image before cutting a fixed grid, so that the blocks are consistent by construction~\citep{zhang2011phase}.
Resolution hierarchies, methods that solve the image first at a coarse scale and then at finer scales, also start from a first solution, in their case a coarse one.
However, those without side-by-side units, that is, units that are neighbors in the image, lie outside this review: examples are multigrid least squares~\citep{pritt1996phase} and multiresolution frequency estimation~\citep{davidson1999multiresolution}.

\paragraph{Coordination during the solve.}
In the second mechanism (row~C of Figure~\ref{fig:architectures}), the units cooperate while they are solved.
Huang et al. (2015) assign one strip to each parallel process, and the processes synchronize the full set of residues after each iteration~\citep{huang2015parallel}.
Dong et al. coordinate the units through their shared voxels instead, a voxel being the three-dimensional counterpart of a pixel.
They use dual decomposition, a technique that splits one optimization problem into overlapping subproblems and then makes their shared variables agree.
In this scheme, each unit holds its own copy of every shared voxel.
Lagrange multipliers, weights that are updated at each iteration, penalize the disagreements between these copies.
The iterations stop when less than 0.1\,\% of the overlap voxels disagree~\citep{dong2016phase}.
Barabadi et al. are an exception, because their units do not cooperate during the solve.
They partition the graph of the problem once, with the library METIS, and solve each patch without the terms that link it to other patches~\citep{barabadi2019dualstage}.
Their coordination comes afterwards: the concatenated solutions initialize the same solver on the whole problem.

\paragraph{Cuts composed before integration.}
In the third mechanism (row~D of Figure~\ref{fig:architectures}), the units are drawn so that no border cuts a cluster of residues (Section~\ref{sec:background}).
The branch cuts can then be decided inside each unit.
In four of the six studies of this mechanism, all units are prepared before one integration of the whole image.
Among the included studies, Yu et al. (2011) give the first partition into tiles that each contain whole residue clusters: they group residues into clusters and draw overlapping tiles around them, before any unwrapping~\citep{yu2011residues}.
This first method does not yet join cuts.
On each overlap, the values already unwrapped by a neighboring tile enter the system of the current tile, as in the first mechanism, and no offset is estimated.
Their later methods return branch cuts instead of phase maps, so Equation~\eqref{eq:offset} does not describe them~\citep{yu2013fast,yu2017largescale}.
The 2013 method links the residues of each tile by a minimum spanning tree and integrates along paths that cross none of these trees.
The 2017 report joins the cuts of all tiles and integrates the image once.

For such cuts, Yu et al. (2017, 2019) state and prove a guarantee on the partition itself: conditions under which the local solution, computed tile by tile, is consistent with the global solution~\citep{yu2017largescale,yu2019largescale}.
Among the included studies, the only other proof of this kind is the $L^1$ proof of their 2013 report~\citep{yu2013fast}.
The 2017 report presents its $L^1$ theorem as a correction of that proof: it relays an unpublished comment by Xu, An and Huang, who found the 2013 proof inaccurate because it ignores the boundaries of the sub-interferograms~\citep{yu2017largescale}.
Their 2011 report makes the corresponding claim for the $L^0$ objective and supports it with examples, without a proof~\citep{yu2011residues}.
The conditions of 2017 rest on balanced residue sets.
A balanced set is a group of residues whose polarities, the signs of the residues, sum to zero.
The conditions also use the envelope distance of a set, the largest distance between two of its residues.
The distance between two sets is the smallest distance between a residue of one and a residue of the other.
The $L^0$ objective counts the differences that depart from the wrapped differences (Section~\ref{sec:background}).
Yu et al. state that minimizing this count amounts to minimizing the total length of the branch cuts~\citep{yu2011residues}.
For this objective, the local solution of each balanced set is consistent with the global one when the distance between any two sets exceeds twice the envelope distance of each of the two~\citep{yu2017largescale}.
For $L^1$, a distance larger than the envelope distance of each of the two suffices, with one exception: the sets balanced by the image border, whose cuts may connect to the border.
The envelope distance of such a set also counts the distance from each residue to the nearest border, and these sets need twice this extended distance.
A relaxed version requires less separation: the factors two and one are divided by a ratio $\alpha\geq1$.
In return, it guarantees consistency only when $\alpha=1$.
Otherwise, it only bounds the total length of the local cuts by $\alpha$ times the global optimum.
The authors present $\alpha$ as a ratio to choose: a larger value allows a finer tiling and gives a looser bound.
A multibaseline extension, which combines several interferograms of the same scene, carries this bound to residues of any integer polarity~\citep{yu2019largescale}.
These guarantees are conditional and specific to the cut objective, so a least-squares or learned local solver would need its own argument.
Lan et al. (2023) avoid computing the cuts~\citep{lan2023clusteranalysis}.
They group the residues into balanced sets and cover each set with its convex hull, the smallest convex polygon that contains it.
They then integrate the rest of the image along a path that crosses no hull, and leave the inside of the hulls without a solution.
They rely on the same guarantee, without proving it again.
Lan and Cui also partition with the convex hulls of residue clusters and rely on the same guarantee without proving it again~\citep{lan2025elaborate}.

\subsection{Class IV: units related by a reference, or kept apart}
\label{sec:classIV}
In class~IV, the units are not related through their seams.

\paragraph{Merging against an external reference.}
In most studies of class~IV, an external reference places each unit.
When each unit is fitted separately, an error in one unit does not propagate to another.
In return, a unit that the reference does not reach stays unresolved.
Liu, Zhao and Jezek give the earliest such merging with a described estimator: each region receives a real constant, fitted against absolute displacements measured by speckle tracking, which matches small patches of two radar images~\citep{liu2007synergistic}.
Libert et al. use an integer instead of a real constant: they add to each region the most frequent integer number of cycles between its unwrapped phase and an absolute phase from split-band interferometry~\citep{libert2017splitband}.
The reference can also be a physical model: Wang et al. (2016) estimate one integer per region jointly with the slip on a fault of known geometry~\citep{wang2016using}.
Yi et al. use a reference plane: they attach each sub-region to it by an integer number of cycles~\citep{yi2025novel}.
In fringe projection, Cui et al. use a projected reference: they seed each region on a projected marker line of absolute phase and discard the regions that the line does not cross~\citep{cui2011reliabilityguided}.
Finally, the reference can be a set of ground stations: Zhao et al. (2026) fix four hydrological units, bounded by levees, roads and built areas, and calibrate each unit on its own water-level gauges~\citep{zhao2026wetland}.

\paragraph{Merging refused on purpose.}
Only Song et al. (2024) never merge their units: they obtain one sub-network of persistent scatterers, points with a stable radar echo, per bridge girder~\citep{song2024toward}.
They refer each girder to its own zero-deformation point and refuse to merge across the expansion joints, because girders on either side of a joint move discontinuously.

\subsection{Borderline cases}
\label{sec:borderline}
A few studies do not fit their class or grouping neatly.
\begin{itemize}
\item Hung and Yamada start from jump-free regions, as in class~II. However, they write one equation per boundary and solve all offsets jointly, so they sit with the other joint solves in grouping G5 of class~I~\citep{hung1998phase}.
\item In the recommended mode of Huang and Lai, the sub-maps appear only after a $\pm2\pi$ rule has been iterated over the whole map~\citep{huang2002innovative}. This rule is local, because it compares each pixel with its four neighbors: it subtracts $2\pi$ from a pixel that lies at least $\pi$ above one of them, or adds $2\pi$ to a pixel that lies at least $\pi$ below. The study applies it to the pixels in parallel. Their regions are therefore drawn after a partial unwrapping, not before any unwrapping as class~II assumes. They stay in class~II because each sub-map then receives one integer, without a local solver.
\item Some studies of class~I use no local solver per unit. The smallest units of Hardie et al., Baldi, Chen et al. (2025) and Blackshire et al. are pixels or jump-free areas~\citep{hardie1998robust,baldi2001twodimensional,chen2025phase,blackshire1999realtime}. Zhong et al. (2011, 2014) run one quality-guided unwrapping over the whole image~\citep{zhong2011combined,zhong2013qualityguided}.
\item Some studies of class~II do unwrap inside their regions~\citep{he2015regiongrowing,liu2016regiongrowing,nie2018fast,gao2022phase,gao2023parallel}.
\item Some studies of class~III still add an offset: in one of their two strategies, Costantini et al. (2012) can add a multiple of $2\pi$ to each block~\citep{costantini2012multiscale}, and Fornaro et al. (2001) fit block constants to a coarse external elevation model~\citep{fornaro2001generation}.
\item Two studies of class~III sit close to the limit of criterion I2. This criterion requires units fixed before the local solves and unwrapped at least partly on their own (Section~\ref{sec:methods}). The mask of Zhang et al. (2009) is partly computed after the branch cuts~\citep{zhang2009new}, and Pepe et al. do not demonstrate a separate solve of their tiles~\citep{pepe2011new}.
\item Two studies outside grouping G17 also leave their merging undescribed. Lan and Cui name a stitch of their regions without describing it, but are placed by their units in class~III~\citep{lan2025elaborate}. Wang et al. (2024) say only that their network tiles are ``merged by inverse operation'', without any offset between tiles, and stay with the network tilings of class~I~\citep{wang2024antinoisedesigned}.
\end{itemize}

\subsection{Issues raised by merging across seams}
\label{sec:across}
Three issues arise when units are compared across their seams: how the common reference is fixed, how the offset across one seam is estimated, and how conflicting relations around a loop are treated.

\paragraph{The common reference.}
The merging fixes only the differences between units, so a merged map is defined only up to one constant.
Some unit or some value must therefore fix this constant, and several studies fix it on one unit.
Strand et al. keep the values of the first block when the image has no mask.
With a mask, they start each connected part of the mask from one full block~\citep{strand1999two}.
Jiang et al. likewise align all tiles on the first tile~\citep{jiang2026wrapcountbased}.
When the tiles do not form one connected set, one reference is needed per set, and Yang and He fix one reference tile in each group of tiles connected by overlaps~\citep{yang2026continuous}.
Moevus and Mignotte also start from one tile: they set the offset of the root tile of their tree to zero, then align each of their results by subtracting its minimum~\citep{moevus2026tilebased}.
At the other extreme, Stetson et al. leave the relative offset of disconnected segments to the user~\citep{stetson1997noiseimmune}.

\paragraph{The offset across one seam.}
The seam estimator decides how much each seam pixel weighs in the offset of the whole unit.
On an overlap, two local solutions are compared at the same pixels.
Across a boundary without overlap, the compared pixels are neighbors, so their difference also contains the physical variation of the phase across the boundary.
Estimating an offset then requires an assumption about this variation, usually that it is small.
Table~\ref{tab:estimators} sorts the estimators.
A mean lets every seam pixel contribute, so it is sensitive to outlying differences, whereas a mode, a median or a vote limits their influence.
For example, SEGUE fixes the offset of a region by a majority vote over its boundary voxels, and applies it only if an acceptance criterion holds~\citep{karsa2019segue}.
Stetson et al. also refuse to decide on doubtful seams: they defer a merge whose mean boundary difference falls near a decision point, and make these merges in a second round~\citep{stetson1997noiseimmune}.
A decision point is a value of this mean at which the choice between two neighboring constants switches, so that noise can flip the choice.
Among the estimators that use every seam pixel, Hardie et al. choose the integer by maximum likelihood under Gaussian noise~\citep{hardie1998robust}.
They show that this choice minimizes the squared distance between two bands of pixels, one on each side of the shared border, after one band is shifted by the integer number of cycles.
This minimum is reached at the rounded mean of the differences between the two bands.
Xiong et al. use another explicit criterion: they choose the integer that minimizes the number of jumps along the boundary~\citep{xiong2008phase}.
At the other extreme, Lo et al. and Huang and Xu rely on a single pixel pair across the seam, so no other seam pixel can outvote it~\citep{lo2002surface,huang2010fast}.

\begin{table}[tbp]
\centering\small
\caption{Seam estimators described in the included studies. The table lists the \nTabEstimators studies whose seam estimator, or at least an applied offset, fits one of its rows; each study appears once, under its main estimator. The other studies are not listed: they set no offset across a seam (as in most of classes~III and~IV), place each unit against a reference, do not report their merging, or use a rule that fits no row.}
\label{tab:estimators}
\begin{tabularx}{\linewidth}{@{}>{\raggedright\arraybackslash}X>{\raggedright\arraybackslash}p{0.36\linewidth}@{}}
\toprule
Estimator & Studies\\
\midrule
Mean of the seam differences, rounded or thresholded to a multiple of $2\pi$\textsuperscript{a} & \citep{charette1996robust,stetson1997noiseimmune,hardie1998robust,strand1999two,schwarz1999hybrid,desvignes2000deroulement,jenkinson2003fast,antonopoulos2015tilebased,he2015regiongrowing,liu2016regiongrowing,nie2018fast,zhong2019hierarchical,gao2022phase,gao2023parallel}\\
Mode, median or vote over the seam\textsuperscript{b,c,d} & \citep{gierloff1987phase,towers1991automatic,arevalilloherraez1996robustsimple,carballo2002hierarchical,meneses2005phaseunwrapping,karsa2019segue,li2021fast,moevus2026tilebased,moevus2026adaptive,jiang2026wrapcountbased,onat2026parallel}\\
Integer chosen by another explicit criterion: absolute deviations, number of jumps or discontinuities, quadratic energy & \citep{baldi2001twodimensional,xiong2008phase,zhong2010fast,hao2017phase,kelany2022quantum}\\
One pixel or one pixel pair & \citep{arevalilloherraez2002robust,lo2002surface,huang2010fast,liu2014threedimensional,lu2014fast,chen2025phase}\\
Fit of a local model across the seam & \citep{dalmaucedeno2008fast,cheng2017close,cheng2018novel}\\
Mean or least-squares fit of the seam differences, kept real & \citep{chang1998tapestry,wu2007phase,suo2008data,lin2011subaperture,wang2013improvement,zhao2014study,zhang2016relative,zhou2017topography,long2022phase,zhao2022high,gourdain2024neural,chen2025stitching,yang2026continuous}\\
Offset applied, its kind or seam statistic not described & \citep{friedlander1996model,joo1997automated,wang1999phaseunwrapping,huang2002innovative,huang2002phase,huang2006phase,huang2007phase,costantini2012multiscale,gao2017twodimensional,nie2018study,xia2024object}\\
\bottomrule
\end{tabularx}

\smallskip
\begin{minipage}{\linewidth}\footnotesize\raggedright
\textit{Notes.} Six of the listed studies solve all their offsets jointly~\citep{carballo2002hierarchical,zhong2010fast,kelany2022quantum,gourdain2024neural,jiang2026wrapcountbased,yang2026continuous}; Gierloff, also coded as merging all units together, does not specify the solve (Appendix~\ref{app:included}).\par
\textsuperscript{a} Stetson et al. add $\pm\pi$ to the multiple of $2\pi$ when the two regions have opposite polarities, a binary label that they give to each region~\citep{stetson1997noiseimmune}.\par
\textsuperscript{b} Carballo and Fieguth take their mode over the four redundant unwrappings of two representative pixels, one per region, not over seam pixels.\par
\textsuperscript{c} The estimator of Towers et al. is written out in the thesis of Judge, a second report of the same study~\citep{judge1992quantitative}.\par
\textsuperscript{d} The medians of Moevus and Mignotte and of Onat and \"Ozkazan\c{c} stay real, without rounding~\citep{moevus2026tilebased,moevus2026adaptive,onat2026parallel}.
\end{minipage}
\end{table}

\begin{figure*}[tbp]
\centering\includegraphics[width=\linewidth]{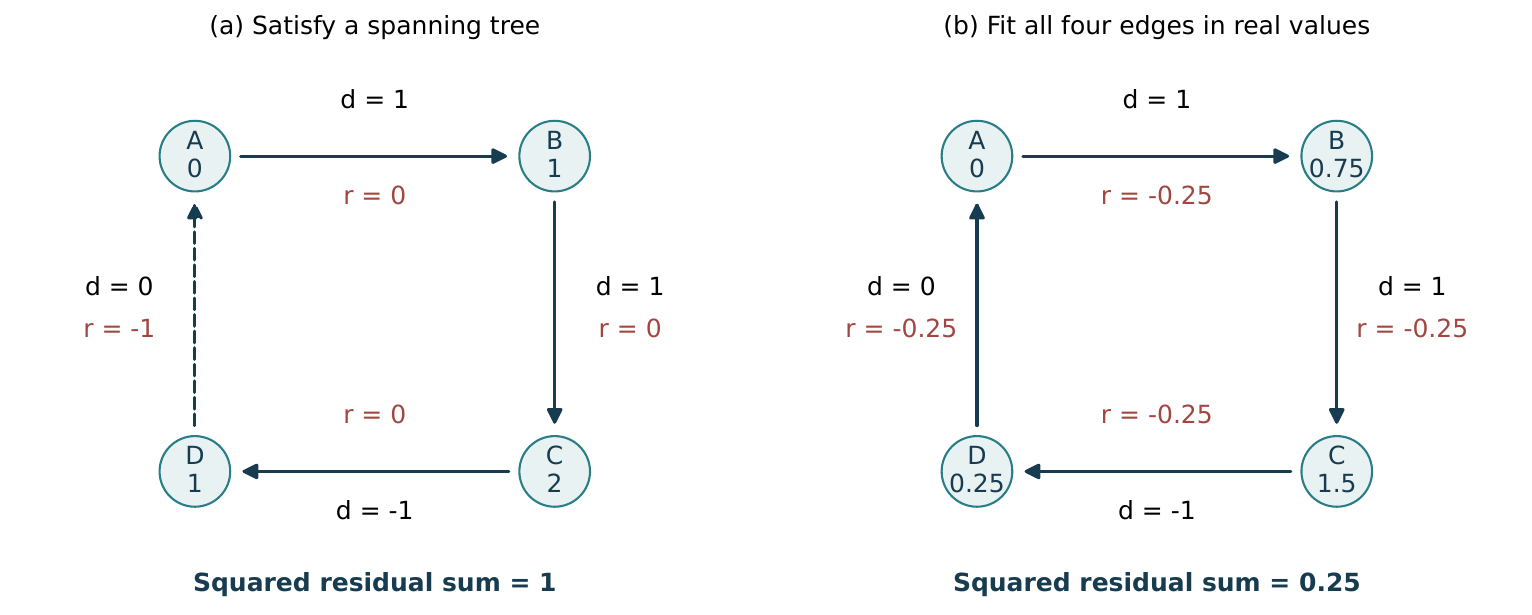}
\caption{\textbf{A tree selects relations; a redundant graph exposes their conflict.} Four constructed boundary observations $d$ sum to one cycle. (a)~The solution satisfies the three relations of a spanning tree and leaves one inconsistent closing edge. (b)~Real least squares spreads the disagreement over the four edges. The residual $r$ of an edge is its offset difference minus $d$. Offsets and observations are in cycles, and $c_A=0$ fixes the common reference.}
\label{fig:cycle}
\end{figure*}

\paragraph{Conflicts around a loop.}
A joint solution and a tree differ when the relations around a loop conflict (Figure~\ref{fig:cycle}).
Consider four units $A,B,C,D$ whose measured differences around a loop are $d=(1,1,-1,0)$ cycles, from $A$ to $B$, $B$ to $C$, $C$ to $D$ and $D$ to $A$.
These differences sum to one cycle, whereas offset differences around a loop always sum to zero.
The residual of a relation is its offset difference minus its measured difference.
Let the offset of $A$ be $c_A=0$, which fixes the common reference.
A spanning tree on the first three relations then gives the offsets $c=(c_A,c_B,c_C,c_D)=(0,1,2,1)$ and leaves a residual of $-1$ on the closing relation, from $D$ to $A$.
Least squares with real-valued offsets gives $c=(0,3/4,3/2,1/4)$ and spreads a residual of $-1/4$ over each relation.
With integer offsets, every residual is an integer.
The four residuals must still sum to $-1$, because the offset differences sum to zero and the measured differences sum to one cycle.
At least one residual is therefore not zero, so the tree solution, with its single residual of $-1$, already minimizes the sum of squared residuals among integer offsets here.
Four integer solutions reach this minimum, one for each relation that carries the residual of $-1$.
The tree decides which one.
The fit residual alone does not establish which solution is closer to the true phase.

The included studies treat the closing relation in several ways.
In a purely sequential or tree order, each unit is fixed once, so the closing relation serves at most as a check.
The grid of Wang et al. (1999) contains such loops, because their blocks overlap in both directions~\citep{wang1999phaseunwrapping,wang1999efficient}.
Their second report joins neighboring border lines sequentially, by a path-dependent method.
Neither report gives the order of the blocks or discusses these loops.
Other studies use the closing relation.
Gierloff already notes that relations between regions can contradict each other around a loop.
In the example of the report, the answer supported by more points of the shared boundaries prevails~\citep{gierloff1987phase}.
Arevalillo Herr\'aez et al. (1996) give the earliest zero-sum test: a group of four zones is masked when the differences around its loop do not sum to zero~\citep{arevalilloherraez1996robustsimple}.
Wu, Feng and Liu need no such test, because they solve each loop of four seams jointly inside their hierarchy~\citep{wu2007phase}.
Towers et al. use the closing relation to explain errors after the fact: they attribute steps at tile edges to two branches of their tree that count different numbers of cycles~\citep{towers1991automatic}.
Chang et al., by contrast, act on a failed loop: they close a loop of merges, and when the loop fails, they declare a singular region, which is divided again or masked~\citep{chang1998tapestry}.
The report calls this failure a closed-loop inconsistency found while stitching, and gives no formula for it.
Jiang et al. act at a larger scale: they reject the whole interferogram when the joint fit is inconsistent (Section~\ref{sec:evidence})~\citep{jiang2026wrapcountbased}.
In their two reports on region residues, Gao et al. (2022, 2023) instead keep the image: they locate the inconsistencies around loops of regions as region residues and pair them by cuts.
The offsets then propagate only along paths that cross no cut~\citep{gao2022phase,gao2023parallel}.

\section{What the reported evidence shows}
\label{sec:evidence}
This section examines what the scores are compared with, how one piece compares with several units, what comparisons exist at a fixed local solver, and how failures and rejected inputs are reported.

\paragraph{What the scores are compared with.}
An error can be counted only against a known target, such as the true phase.
Here, a ground truth is a phase known independently of the method (Section~\ref{sec:methods}).
A ground truth enters at least one evaluation in \nGroundTruth of the \nStudiesIncluded studies, and real data appear in \nRealData.
Both items are coded per study, so one study can score a synthetic case against its true phase and assess its real data visually.
Hardie et al.\ do so across their two reports~\citep{hardie1998robust,blackshire1999advanced}, and Fornaro and Sansosti within one report~\citep{fornaro1999twodimensional}.

Other comparisons measure properties other than an error against a known phase.
A rewrapping residual tests congruence, not the recovery of the right cycles: Wang et al.\ (2004) report that 2.6\,\% of pixels differ after rewrapping a simulation~\citep{wang2004phase}.
Agreement between two methods is not an error against a known phase either: Costantini et al.\ find that block decomposition and their multiscale solution unwrap 0.19 to 2.12\,\% of pixels differently~\citep{costantini2012multiscale}.
Pepe et al.\ use a third kind of measure, a consistency criterion over their series of interferograms, and by this criterion they judge more than 92\,\% of pixels correctly unwrapped~\citep{pepe2011new}.
The same study also compares its deformation time series with leveling and GPS measurements, which is its ground truth.

An independent measurement counts as a ground truth here, but it carries its own limits.
Xie et al.\ (2010) score their results against images acquired at a higher velocity encoding, where the phase wraps less, and treat these images as ground truth~\citep{xie2010patchbased}.
Li et al.\ (2015) check their subregions against a single GPS station, so their check rests on one point: the displacement along the radar's line of sight is 20.0\,cm with subregions, 8.3\,cm with SNAPHU in one piece and 17.1\,cm from GPS~\citep{li2015coseismic}.
A reference that also calibrates the result, by contrast, is not independent.
Zhao et al.\ (2026) report a root-mean-square error (RMSE) of 4.03\,cm with hydrological units, against 13.20\,cm in one piece, measured at water-level gauges~\citep{zhao2026wetland}.
Their offsets are calibrated on the same gauges, so this RMSE is not an independent check.

\paragraph{One piece against several units.}
Running the same local solver on the same image, once in one piece and once on the units, measures the partition and the merging together, not either one alone.
This comparison appears in \nOneVsMany of the \nStudiesIncluded studies.
As shown below, only \nOVMNumberTruth of these studies give a number for the one-piece comparison itself against a known phase.
A comparison against other methods does not isolate even this joint effect, because the solver and the merging rule change at the same time.

These \nOneVsMany comparisons do not all measure the same thing.
Of the \nOneVsMany studies, \nOVMTimeOnly report only time or memory, and \nOVMVisual compare the accuracy only on images.
The other \nOVMNumber give a number, but this number is not always an error.
Among these \nOVMNumber studies, \nOVMNumberTruth measure an error against a known phase, and the others measure the agreement between the two runs or another criterion, such as the length of the branch cuts, an objective value or a closure error.
A ground truth elsewhere in the report does not change this count.
Table~\ref{tab:onevsmany} gives these counts.
In a separate count, \nGTandOVM of the \nOneVsMany studies use a ground truth in at least one evaluation, but not always for the one-piece comparison.
Huang and He illustrate the comparisons of time alone: they report 3\,min 50\,s instead of an hour and a half on an optical interferometry map, and only state that the result stays satisfactory~\citep{huang2002phase}.
Lo et al.\ illustrate the visual comparisons: they reduce about 90\,min to about 130\,s, and judge on the images that the result is the same~\citep{lo2002surface}.
Huang et al.\ (2015) and Yu et al.\ (2011) also compare their tiled and one-piece results visually~\citep{huang2015parallel,yu2011residues}.

\begin{table}[tbp]
\centering\small
\caption{What the \nOneVsMany comparisons of one piece against several units report. The last row is a subset of the row above it.}
\label{tab:onevsmany}
\begin{tabular}{@{}lr@{}}
\toprule
What the comparison reports & Studies\\
\midrule
Time or memory only & \nOVMTimeOnly\\
Accuracy compared on images only & \nOVMVisual\\
A number & \nOVMNumber\\
\quad of which an error against a known phase & \nOVMNumberTruth\\
\bottomrule
\end{tabular}
\end{table}

Beyond saving time, partitioning can also make a computation possible.
Chen and Zebker cut the memory per processor from 300 to 25\,MB on a $2000\times1760$ extract with $4\times4$ tiles~\citep{chen2002phase}.
Yu et al.\ (2017) also cut the peak memory, from 7600.67\,MB in one piece to 68.80\,MB on a $3000\times3000$ image~\citep{yu2017largescale}.
At a larger scale, Zhang et al.\ (2011) unwrap a satellite radar scene of $120{,}000\times9274$ pixels in 17\,min 14\,s on 8 cores, where one-piece minimum-cost flow would have needed more than 250\,GB of memory~\citep{zhang2011phase}.
Merging can still dominate the runtime: on a sonar interferogram, in the parallel run of Zhong et al.\ (2019), the merge on the main node lasts longer than the unwrapping of the blocks~\citep{zhong2019combined}.

Where a number is given, several studies report a small cost, or none, although they measure different things.
Yu et al.\ (2017) measure the length of the branch cuts: against the same $L^1$ solver in one piece, their partition lengthens these cuts by at most 1.32\,\% on an interferogram of the Etna volcano~\citep{yu2017largescale}.
In the multibaseline extension, Yu et al.\ (2019) compare objective values instead: the sum of the local optima equals the global optimum on simulated images of up to $1400\times1400$ pixels, and exceeds it by 7 out of 531{,}562 at $3000\times3000$~\citep{yu2019largescale}.
Dong et al.\ measure the agreement between the two runs: in MRI, they compute up to 18 times faster than the same graph cut in one piece, with at most 28 voxels in disagreement~\citep{dong2016phase}.
Zhong, Tang and Zhang count 45{,}765 pixels where the result is discontinuous after merging, against 45{,}703 in one piece, in less than a quarter of the time with several threads~\citep{zhong2012blocked}.
Of these four studies, only Zhong, Tang and Zhang also score the one-piece comparison against a true phase.
On the Long's Peak interferogram, the variance of their error, given without a unit, rises from 1.533 to 1.994 over the whole image, and stays near 0.30 in high-quality areas.

Other studies report a cost in accuracy.
Chen and Zebker judge it by eye: on maps of the error against a phase simulated from an elevation model, which they call coarse, the one-piece solution of their extract contains fewer errors than the tiled solution~\citep{chen2002phase}.
Wu, Feng and Liu give a number instead.
On a $256\times256$ simulation, they report an error, which they do not define, of 0.3507\,rad in one piece, against 1.9322\,rad with $128\times128$ blocks and 2.2186\,rad with $64\times64$ blocks, while the time falls from 8710\,s to 2043\,s and 285\,s~\citep{wu2007phase}.
With their region-division method, Nie et al.\ (2018) find the same trade-off between time and errors: they take 4.5\,s instead of 331.7\,s for quality-guided unwrapping in one piece, but count 100 erroneous pixels against 53~\citep{nie2018fast}.
Jiang et al.\ measure the cost with a closure error rate, computed over triangles of three interferograms and not over loops of units.
This rate is the fraction of unmasked pixels whose phase, summed around such a triangle, misses zero by more than $\pi$\,rad, counting only connected areas of at least 100 pixels.
With partitioned minimum-cost flow, the mean rate is 5.4\,\%, against 1.1\,\% for the same solver in one piece; the partitioned mode of SNAPHU gives 3.6\,\%, with no one-piece value~\citep{jiang2026wrapcountbased}.

In the other direction, Haghighi and Dimopoulos report a gain in accuracy.
They obtain a mean agreement of 86\,\% with a partition, against 70\,\% for the same solver in one piece, on five $100\times100$ images~\citep{haghighi2024enhanced}.
Here, agreement is the fraction of pixels identical to a noisy unwrapped ground-truth image.

Some of these comparisons change more than the partition, so their outcome cannot be attributed to the partition alone.
Xu et al.\ reduce the unwrapping time from 324.389\,s to 2.654\,s, but the filtering and the removal of the flat-Earth phase, the phase ramp that a terrain without relief would produce, also change~\citep{xu2015phase}.
Yi et al.\ report a residual standard deviation of 0.087\,rad with sub-regions, against about 3.39\,rad for two quality-guided algorithms on the whole image, but their sub-regions are also tied to an external reference plane~\citep{yi2025novel}.
The comparison of Song et al.\ (2026) is not counted among the \nOneVsMany: they report a normalized RMSE of 0.46\,\% with tiles, against 0.68\,\% for the whole image, but their one-piece run resizes the image to the network input, so it also changes the resolution~\citep{song2026insar}.

Across these comparisons, the outcome depends on the method, the data and the measure, so they do not add up to a general effect of partitioning.

\paragraph{Comparisons at a fixed local solver.}
Only \nMergeOrderPairs studies compare merging orders as such, on tiles unwrapped by the same local solver~\citep{antonopoulos2015tilebased,chen2025stitching}.
Antonopoulos et al.\ compare a unidirectional merge with their reliability-ordered merge, on tiles unwrapped by Strand's method~\citep{antonopoulos2015tilebased}.
In a second contrast, which also changes the tile solver, their polynomial tiles merged by reliability alter only the four tiles that contain the residue of a synthetic map, whereas pixel-level SRNCP and Strand's method spread the error over about a third and a quarter of the image.
Chen, Wang and Zhang report simulated mean absolute errors of 1.8179\,rad for least squares solved with the discrete cosine transform (DCT) in one piece, 1.2325\,rad for blocks merged in a serpentine order and 0.2201\,rad for blocks merged in their quality-guided order~\citep{chen2025stitching}.
Their report does not state that the block size and the offset estimator are the same in the serpentine variant.

Other studies compare merging rules, sometimes together with the order.
Onat and \"Ozkazan\c{c} compare three merging methods on the same tiles, a unidirectional merge, a reliability-ordered merge and their own median-based rule, which differ in both the order and the rule; their own gives the lowest RMSE of the three~\citep{onat2026parallel}.
Even with this rule, the RMSE generally grows with the number of tiles, on all four datasets.
Three other studies also compare merging rules, and a fourth, by Jiang et al., is described below with the rejected inputs.
Wang et al.\ (2013) replace the rounded multiple of $2\pi$ of an earlier tiled method with a real mean of the differences on the overlaps, together with other changes to the merging; on a topographic map, the mean relative error falls from 1.76\,\% to 1.28\,\%, but the report does not define this error~\citep{wang2013improvement}.
On a simulated field whose tiles receive injected scales and offsets, Yang and He solve one scale and one offset per tile jointly, and reach a peak signal-to-noise ratio of 27.9\,dB, against 20.6 to 22.8\,dB for three blends without correction~\citep{yang2026continuous}.
Xie et al.\ (2021) compare ways of finding the lateral shift between subapertures: on 50 datasets, their combined matching succeeds in 80\,\% of the cases, against 28 to 48\,\% for each single method~\citep{xie2021phase}.

Among the included studies and under our coding, one study, a preprint by the authors of this review, measures how the partition construction changes runtime and accuracy, with the local solver and the merging rule held fixed~\citep{moevus2026adaptive}.
It compares a regular grid with quadtree and k-d tree partitions that split the tiles containing residues; all use DCT least squares and merge the tiles by the median of the seam differences, along a reliability tree.
On 198 inputs, the adaptive partitions are 23 to 58\,\% slower than the grid (median of runtime ratios), and the mean S$_\pi$, the fraction of pixels whose error is below $\pi$ after a common offset is removed, drops by 0.04 to 1.02 percentage points.
In the quadtree, building the partition takes about a quarter of the total time.
On two related test images of a shear field, S$_\pi$ falls from 100\,\% with the regular grid to 0\,\% with the adaptive partitions.

Other studies vary the size or the number of tiles, with the local solver and the merging rule fixed within each series, and often vary other factors too, such as the overlap, the number of threads or the solver.
Some measure only the cost, as time and sometimes memory~\citep{huang2002phase,dudczyk2014optimizing,barabadi2019dualstage}.
Sawaf and Tatam measure even less: they time only the solver of one tile, on random-weight graphs of 64 to 2048 pixels per side, without a phase map~\citep{sawaf2006finding}.
Others measure accuracy or a quality indicator, with or without time~\citep{wu2007phase,kelany2022quantum,dong2016phase,gao2017twodimensional,onat2026parallel}.
For example, on a real image without a reference, Gao et al.\ (2017) report a shorter run with $14\times6$ blocks than with $7\times3$ (7.6\,s against 18.0\,s), for a nearly equal $L^0$ norm (13{,}729 against 13{,}714 pixels), which counts the departures of the result from the wrapped differences (Section~\ref{sec:background}).
Huang et al.\ (2006) also contrast two constructions, a static and a dynamic segmentation, but give only the count of thin border regions that each leaves, 71 and 5, and state without a score that the static segmentation alone gives an incorrect map~\citep{huang2006phase}.

\paragraph{Failures and rejected inputs.}
A score computed after rejecting inputs describes only the retained set.
Jiang et al.\ show this case when they merge minimum-cost-flow tiles with their own joint rule instead of the rule of the partitioned solver~\citep{jiang2026wrapcountbased}.
This rule rejects 137 of 846 interferograms as inconsistent.
On the 3074 triangles formed by the interferograms that it keeps, the mean closure error rate is 1.6\,\%, against 3.5\,\% for the same tiles merged by the partitioned solver.
These two rates cannot be compared with the 5.4\,\% given above, which covers all triangles of the dataset.

A selection of points can also change the result.
Libert et al.\ correct 33 to 74 of 1796 regions of their radar image pair, depending on the criterion that selects the points they use, and they leave a region uncorrected when it has fewer than 10 of these points~\citep{libert2017splitband}.
Stephenson et al.\ make the selection of tiles explicit: they classify each tile for unwrapping, repair, acceptance as is, a second unwrapping at a lower tolerance or rejection~\citep{stephenson1994data}.

Reported failures are also part of the evidence.
Baldi reports a failure on a synthetic shear along the diagonal~\citep{baldi2001twodimensional}.
Liu and Drangova report that stacking slices removes the errors between slices in the brain but not in the tongue or the neck, where a second step is needed~\citep{liu2012interventionbased}.

\section{Discussion}
\label{sec:discussion}

\subsection{Main findings}
This review has three main findings.
First, building the units and merging them are two distinct decisions.
The two decisions are linked by what each unit returns, that is, its local output.
For example, on a fixed grid, a unit can return its block unwrapped by a constant phase rotation, a polynomial corrected by the wrapped residual, or a block model used only as a reference for rounding (Section~\ref{sec:classI}).
Two methods that share a partition can therefore pose different merging problems.

Second, the values allowed for the merging variable decide what the merged field guarantees.
Integer offsets applied to congruent local outputs keep the merged field congruent.
By contrast, real-valued offsets generally do not keep it congruent.
Boundary conditions imposed on least-squares solves generally do not either.
How the values are obtained matters as well.
A joint solve exposes the conflicts between seam relations.
A purely sequential or tree order, by contrast, fixes each unit once, so the relation that closes a loop of units serves at most as a check (Section~\ref{sec:across}).

Third, the evidence rarely isolates either decision.
The most common test runs the same local solver on the same image, once in one piece and once on the units.
It appears in \nOneVsMany studies.
These comparisons, however, change the partition and the merging together.
In addition, \nOVMTimeOrVisual of them report only time or memory, or assess accuracy visually.
Tests of one decision at a time are rarer.
For the merging, few studies compare orders or rules at a fixed local solver~\citep{antonopoulos2015tilebased,chen2025stitching,wang2013improvement,xie2021phase,onat2026parallel,jiang2026wrapcountbased,yang2026continuous}.
For the partition, among the included studies and under our coding, one study measures how the construction changes runtime and accuracy, with the local solver and the merging rule held fixed~\citep{moevus2026adaptive}.
A few others vary the size or the number of tiles, and often vary other factors too (Section~\ref{sec:evidence}).
On the theoretical side, a guarantee on the partition itself is proved only in the reports of Yu et al., under conditions specific to the objective that is minimized~\citep{yu2017largescale,yu2019largescale}.
These proofs follow a first proof that the same authors later corrected~\citep{yu2013fast}.

\subsection{Implications for design and evaluation}
Based on this review, we offer some suggestions.
They are our own, not rules drawn from the included studies.
The first suggestion is to judge a partition against the purpose it serves.
For example, to support a memory claim, a study can report the peak memory of the local solves and the merging.
Likewise, to support a speed claim, it can report the total time, including the construction of the partition and the merging.
Finally, to support an accuracy claim, it can report the quantity that the result is compared with, such as a known phase, and how the constant offset between the two was removed.
Table~\ref{tab:reporting} lists these items and the others worth stating, so that partitioned methods can be compared.

\begin{table*}[tbp]
\centering\small
\caption{Our suggestions for reporting a partitioned unwrapping method. The last column gives examples from the included studies. Local model: a simple function, such as a polynomial, fitted to the phase of one unit.}
\label{tab:reporting}
\begin{tabularx}{\linewidth}{@{}>{\raggedright\arraybackslash}p{0.2\linewidth}>{\raggedright\arraybackslash}X>{\raggedright\arraybackslash}X@{}}
\toprule
Item & What to state & Examples in the included studies\\
\midrule
Purpose & Memory, time, local model, acquisition or physical boundary, and the outcome measured for it & Memory reported against one piece~\citep{chen2002phase,yu2017largescale}; tile size judged on time alone~\citep{dudczyk2014optimizing}\\
Construction & Geometry, criterion and moment at which the units are fixed & A binary tree of blocks fixed in advance~\citep{hardie1998robust}; regions drawn on the wrapped phase~\citep{gierloff1987phase}; a rupture trace~\citep{li2015coseismic}\\
Local output & Phase map, integer labels, flows or branch cuts & Cuts composed before one integration~\citep{yu2017largescale}; flows fixed on overlaps~\citep{costantini1998novel}\\
Merging variable & Its dimension and domain: integer, real, scale or geometric parameters & A real scale and offset per tile~\citep{yang2026continuous}; piston, tilts and defocus~\citep{zhao2014study}\\
Estimator and order & Seam statistic, order or joint solve, common reference & Table~\ref{tab:estimators}; reliability tree~\citep{towers1991automatic}; joint least squares~\citep{hung1998phase}\\
Loops & Whether redundant relations are kept and how conflicts are treated & A loop closed as a check~\citep{chang1998tapestry}; inconsistent inputs rejected~\citep{jiang2026wrapcountbased}\\
What the score is compared with & Ground truth, rewrapping residual, other method or external measurement, with its limits & A single GPS station~\citep{li2015coseismic}; a higher velocity encoding~\citep{xie2010patchbased}\\
Retained inputs & Rejected inputs and units left unresolved & Interferograms rejected~\citep{jiang2026wrapcountbased}; regions left uncorrected~\citep{libert2017splitband}\\
Cost & Construction, local solves and merging, separately & Construction a quarter of the time~\citep{moevus2026adaptive}; on sonar data, merge on the main node longer than the parallel solves~\citep{zhong2019combined}\\
\bottomrule
\end{tabularx}
\end{table*}

The second suggestion goes beyond reporting: three experiments would address the gaps in the evidence.
First, with the local solver and the merging rule fixed, one can vary the partition and report the construction cost, the total cost and an explicitly stated accuracy measure.
Second, with the local outputs fixed, one can vary the merging rule and report the seam disagreements, the conflicts around loops, the final rewrapping residual and, when the true phase is known, the final accuracy.
Third, where feasible, one can run the complete method in one piece and on several units, with the same inputs, masks and tolerances, and report an explicitly stated accuracy measure rather than time alone.
This third experiment measures the joint effect of partitioning and merging, which the first two separate.
Each of these experiments is informative even when it shows no benefit.

\subsection{Limitations}
Studies counted separately can share authors, methods or data.
Our verification also has a limited scope.
A verified statement shows that a report says what this review attributes to it.
It does not show that the reported result is correct.

The review process has further limits.
The search used English terms in titles and abstracts, so reports in other languages, or reports that describe a partition in other words, may have been missed.
The citation search followed the references and the citing works of \nSeeds seed reports, in OpenAlex only and over one generation only.
The \nCitNoAbstract citation candidates without a published abstract were screened on their title alone.
The partition terms of the queries contain no word specific to volumes, such as subvolume, so three-dimensional methods may be underrepresented.
The records excluded only for being three-dimensional were, however, screened again.
Five reports could not be obtained.
Four of them have no full article, and the revised version of the fifth was read.
For the four others, the available information does not show whether they describe another construction or merging rule.
The second author independently screened one fifth of the database records but none of the citation candidates, and read one fifth of the full texts.
At screening, the two authors disagreed on \nDisagreeScreening of the \nScreenSample sampled records, so the first author's screening alone may have excluded eligible records.
Neither of the two included preprints by the authors fell in the samples read by the second author.
The criteria were fixed before the reading, except the admission of three-dimensional methods on 29 October 2026.
The way they were applied to borderline cases was written down at the end of it.
The reading focused on the method sections.
No extraction was duplicated by the second author, and no risk-of-bias tool or certainty grading was applied.
Reporting bias was not assessed, and the review was not registered.
The rules for coding a ground truth and a one-piece comparison were refined after the reading, before the final count (see Data availability).
Finally, the groupings were proposed during extraction, and the first author validated their placements.
Another reader could therefore draw some boundaries differently.
For \nAxesAmbiguous studies, two codes could be defended for the merging.
With the alternative codes, no count of the merging variable or of the estimation procedure changes by more than \nSensMaxShift studies, except the studies that adjust another variable, which go from \nVarOther to \nSensOtherAlt.

\section{Conclusion}
\label{sec:conclusion}
Partitioned phase unwrapping rests on two decisions: how the units are built and how their local solutions are merged.
What each unit returns links the two decisions.
Together with the values allowed for the merging variable, it also fixes what the merged field can guarantee.
Across the \nStudiesIncluded included studies, regions computed before unwrapping and fixed tiles are the most frequent constructions, and one integer offset per unit is the most frequent merging variable.
Four classes, set by where the relation between units is decided, organize these studies (Figure~\ref{fig:hierarchy}).
The classes are the following: assembling locally unwrapped maps (class~I), one integer per jump-free region (class~II), consistency built into the solve (class~III), and units related by a reference, or kept apart (class~IV).
The \nGroupNR studies outside these classes (grouping G17) describe their partition but not their merging.

Overall, the reported evidence rarely isolates either decision.
The reported scores also measure different properties, such as congruence, agreement with another method or accuracy against a known target, so they cannot be pooled.

\section*{Declarations}
\paragraph{Competing interests.}
The authors wrote two of the included studies, both preprints~\citep{moevus2026tilebased,moevus2026adaptive}.
The OpenAlex search found them, and they followed the same procedure as every other report (Section~\ref{sec:methods}).
Neither fell in the random samples read by the second author, at screening or on the full text.
The authors declare no other competing interests.

\paragraph{Funding.}
This research did not receive any specific grant from funding agencies in the public, commercial, or not-for-profit sectors.

\paragraph{Registration and protocol.}
The review was not registered in a public protocol registry.
Instead, a dated protocol fixed the eligibility criteria on 1 October 2026, before the first author's screening, with later dated amendments, and the rules for applying them on 29 October 2026.

\paragraph{Data availability.}
The seed reports of the citation search, the included reports with the study each belongs to, both merging codes of each study, the protocol and the other registers are available from the authors on request.
The protocol and the registers are written in French.

\paragraph{CRediT authorship contribution statement.}
Antoine Moevus: Conceptualization, Methodology, Investigation, Data curation, Validation, Visualization, Writing -- original draft.
Max Mignotte: Conceptualization, Supervision, Validation, Writing -- review and editing.

\paragraph{Declaration of generative AI and AI-assisted technologies in the manuscript preparation process.}
The first author designed the protocol, the eligibility criteria and the search, and screened every record.
The first author read every full text, focusing on the method sections, and decided its eligibility.
The first author also validated the included studies and their groupings.
During the preparation of this work, the authors used Claude Code, a commercial AI agent, to assist the exploration and refinement of the search queries and the removal of duplicates.
It was also used to facilitate the first screening of the titles and abstracts, to keep the registers up to date, to prepare the extraction table and to make some verifications against the full texts.
Finally, it was used to correct the spelling.
After using this tool, the authors reviewed and edited the content as needed and take full responsibility for the content of the published article.

\bibliographystyle{unsrtnat}
\bibliography{references}
\clearpage
\onecolumn
\appendix
\counterwithin*{table}{section}\renewcommand{\thetable}{\thesection.\arabic{table}}
\section{Search strings and exclusion codes}
\label{app:search}
\paragraph{Queries.}
Each query joins concept A, phase unwrapping, to concept B, partition and merging, in titles and abstracts (Table~\ref{tab:queries}).
Concept A reads \texttt{"phase unwrapping" OR "unwrapped phase"}.
No date or document-type filter was applied.
The OpenAlex queries used the \texttt{title\_and\_abstract.search} field.
Concept B lists each inflected form, with hyphenated and joined spellings.
The main OpenAlex query reads:

{\small\raggedright\ttfamily ("phase unwrapping" OR "unwrapped phase") AND (tile OR tiles OR tiled OR tiling OR tilings OR block OR blocks OR blockwise OR "block-wise" OR "block-based" OR subblock OR subblocks OR "sub-block" OR "sub-blocks" OR subimage OR subimages OR "sub-image" OR "sub-images" OR subregion OR subregions OR "sub-region" OR "sub-regions" OR subdomain OR subdomains OR "sub-domain" OR "sub-domains" OR subinterferogram OR subinterferograms OR "sub-interferogram" OR "sub-interferograms" OR partition OR partitions OR partitioned OR partitioning OR quadtree OR quadtrees OR "quad-tree" OR "quad-trees" OR "divide and conquer" OR "divide-and-conquer" OR mosaic OR mosaics OR mosaicking OR mosaicing OR stitch OR stitches OR stitched OR stitching OR merge OR merges OR merged OR merging OR patch OR patches OR patched OR patching OR piecewise OR "large-scale" OR "large scale")\par}

\noindent A complementary query joins concept A to \texttt{subproblem OR subproblems OR "sub-problem" OR "sub-problems"}.
Below, \texttt{A} stands for the first parenthesis of the main query, without its parentheses, and \texttt{B} for the second.
Web of Science used \texttt{(TI=(A) OR AB=(A)) AND (TI=(B) OR AB=(B))}, then the same string with the four complementary terms in place of \texttt{B}.
Scopus used \texttt{TITLE-ABS(A) AND TITLE-ABS(B)}, with the four complementary terms added at the end of \texttt{B}.
IEEE Xplore limits the number of terms in a query, and its search already matches plural forms, so it received five queries on singular terms.
The first reads:

{\small\raggedright\ttfamily ("Document Title":"phase unwrapping" OR "Document Title":"unwrapped phase" OR "Abstract":"phase unwrapping" OR "Abstract":"unwrapped phase") AND ("Document Title":tile OR "Document Title":block OR "Document Title":subblock OR "Document Title":"sub-block" OR "Document Title":subimage OR "Document Title":"sub-image" OR "Abstract":tile OR "Abstract":block OR "Abstract":subblock OR "Abstract":"sub-block" OR "Abstract":subimage OR "Abstract":"sub-image")\par}

\noindent The four others keep the first parenthesis and replace the terms of the second, in both fields.
The five groups of terms are: \texttt{tile, block, subblock, sub-block, subimage, sub-image}; \texttt{subregion, sub-region, subdomain, sub-domain, subinterferogram, sub-interferogram}; \texttt{partition, quadtree, quad-tree, divide and conquer, mosaic, mosaicking}; \texttt{stitch, merge, patch, piecewise, large scale}; \texttt{subproblem, sub-problem}.
A pilot rejected the terms \texttt{grid}, \texttt{decompose}, \texttt{divided into}, \texttt{split into}, \texttt{domain decomposition} and \texttt{region division}, whose added records were mostly off topic.

\begin{table}[!htbp]
\centering\small
\caption{Databases searched on 1 October 2026.}
\label{tab:queries}
\begin{tabularx}{\linewidth}{@{}>{\raggedright\arraybackslash}X>{\raggedright\arraybackslash}p{0.42\linewidth}r@{}}
\toprule
Database & Fields and queries & Records\\
\midrule
OpenAlex & title and abstract; main and complementary queries & \nOpenAlex\\
IEEE Xplore & Document Title and Abstract; one query per term group & \nIEEE\\
Web of Science Core Collection & TI and AB; main and complementary queries & \nWoS\\
Scopus & TITLE-ABS; one query & \nScopus\\
\midrule
Total & & \nRecordsRaw\\
\bottomrule
\end{tabularx}
\end{table}

\paragraph{Exclusion codes.}
Table~\ref{tab:exclusions} lists the exclusion codes, assigned by the first author, at both stages.

\begin{table*}[!htbp]
\centering\small
\caption{Exclusion codes and counts. The two screening columns count records, from the databases and from the citation search; the full-text column counts reports from all sources, other methods included. Each excluded item has one reason. A code defined in the protocol but never used keeps its row, with zeros.}
\label{tab:exclusions}
\begin{tabularx}{\linewidth}{@{}>{\raggedright\arraybackslash}X r r r@{}}
\toprule
Reason & Screening, databases & Screening, citations & Full text\\
\midrule
No partition fixed before the local solves (global solve, mask, ordering of a single path, or blocks used only to estimate a parameter) & \nScrExclSansPartition & \nCitScrExclSansPartition & \nExclSansPartition\\
Units created during unwrapping (region growing, flood fill) & \nScrExclRegionsPendant & \nCitScrExclRegionsPendant & \nExclRegionsPendant\\
Non-spatial decomposition (frequency, time, wavelength or radar baselines) & \nScrExclNonSpatial & \nCitScrExclNonSpatial & \nExclNonSpatial\\
Units never merged into one phase field (separate objects, or networks kept apart) & \nScrExclObjetsSepares & \nCitScrExclObjetsSepares & \nExclObjetsSepares\\
Resolution hierarchy without side-by-side units & \nScrExclHierarchie & \nCitScrExclHierarchie & \nExclHierarchie\\
Patches used only to train a network & \nScrExclPatchsEntrainement & \nCitScrExclPatchsEntrainement & \nExclPatchsEntrainement\\
One-dimensional unwrapping & \nScrExclUneDimension & \nCitScrExclUneDimension & \nExclUneDimension\\
No phase unwrapping, off topic, or a partition proposed without any unwrapping & \nScrExclHorsSujet & \nCitScrExclHorsSujet & \nExclHorsSujet\\
Application of an existing unwrapper without a partitioned method & \nScrExclApplication & \nCitScrExclApplication & \nExclApplication\\
General review & \nScrExclRevueGenerale & \nCitScrExclRevueGenerale & \nExclRevueGenerale\\
Method not described well enough to judge the partition & \nScrExclPasUnArticle & \nCitScrExclPasUnArticle & \nExclPasUnArticle\\
Retracted by the publisher & \nScrExclRetracte & \nCitScrExclRetracte & \nExclRetracte\\
\midrule
Total & \nExcludedScreening & \nCitExcluded & \nExcludedFullText\\
\bottomrule
\end{tabularx}
\end{table*}

\section{Included studies}\label{app:included}
Table~\ref{tab:included} lists the \nStudiesIncluded included studies by grouping. Each study appears once, under its main report. Another report of the same study that this review cites follows ``also''; the other reports, which are conference or earlier versions of a listed report, are not listed here (see Data availability). Modality, merging type and the three evidence columns come from the extraction verified against the full texts. ``GT'' marks an evaluation against a phase known independently of the method (simulation, calibrated object or independent measurement), ``1 vs $n$'' a comparison of the same local solver on the same image, run in one piece and on the units, and ``Measured'' the use of real data, acquired by an instrument rather than simulated (Section~\ref{sec:methods}). Table~\ref{tab:groupings} first gives the key to the grouping numbers of Figure~\ref{fig:hierarchy}.
\begin{table*}[!htbp]\centering\small
\caption{The \nGroups groupings of the included studies, by mechanism. Most groupings combine how the units are built with how they are merged; G6 and G7 are defined by the construction alone, G13 and G16 by the merging alone. QUBO: quadratic unconstrained binary optimization; CQM: constrained quadratic model; PRELUDE and Gierloff type: see Section~\ref{sec:classII}. The numbers refer to Figure~\ref{fig:hierarchy}. The last column gives the earliest included study of each grouping; Appendix~\ref{app:included} lists all studies.}\label{tab:groupings}
\begin{tabularx}{\linewidth}{@{}>{\raggedright\arraybackslash}Xr>{\raggedright\arraybackslash}p{0.26\linewidth}@{}}
\toprule Grouping & Studies & Earliest included study\\\midrule
G1 Fixed grid, one integer offset per tile, set sequentially or along a reliability tree & 12 & \citet{towers1991automatic}, 1991\\
G2 Fixed grid, one real-valued offset per tile (seam mean or median, sign and constant), sequential or quality-guided & 7 & \citet{chang1998tapestry}, 1998\\
G3 Adaptive or hierarchical partitions (binary tree, quadtree, k-d tree, content-shifted blocks) & 9 & \citet{arevalilloherraez1996robustsimple}, 1996\\
G4 Regions unwrapped by a local solver, then joined by an integer or real-valued offset & 8 & \citet{ching1992two}, 1992\\
G5 All offsets solved jointly: secondary network, least squares on a graph, QUBO or CQM, discontinuity minimization & 16 & \citet{hung1998phase}, 1998\\
G6 Partition imposed by acquisition (subapertures, slices, frames, scenes) & 11 & \citet{ramesh2000phase}, 2000\\
G7 Tiling imposed by the fixed input size of a neural network & 5 & \citet{zhao2021phase}, 2021\\
G8 Jump-free regions from the wrapped phase, one integer per region, greedy merging or propagation (PRELUDE lineage and related methods) & 19 & \citet{stetson1997noiseimmune}, 1997\\
G9 Jump-free fringe regions, one integer per region (Gierloff type) & 14 & \citet{gierloff1987phase}, 1987\\
G10 Integer region labels chosen by an energy on the region graph (graph cut, Markov random field) & 4 & \citet{xie2010patchbased}, 2010\\
G11 Blocks with imposed values, or a coarse global solution computed before partitioning, mostly without offset estimation & 5 & \citet{costantini1998novel}, 1998\\
G12 Reliability regions unwrapped in order, each anchored by Dirichlet conditions or constraints from the area already unwrapped & 9 & \citet{fornaro1999twodimensional}, 1999\\
G13 Coordination during solving or exact resolution: no merging afterwards (synchronized strips, warm start, dual decomposition) & 3 & \citet{huang2015parallel}, 2015\\
G14 Partitions that cut no residue cluster: local branch cuts joined, then a single integration & 6 & \citet{yu2011residues}, 2011\\
G15 Units tied to an external absolute reference, without a shared seam & 8 & \citet{liu2007synergistic}, 2007\\
G16 Merging deliberately refused because the discontinuity is physical & 1 & \citet{song2024toward}, 2024\\
G17 Partition described, merging not described or deferred to another source & 12 & \citet{stephenson1994data}, 1994\\
\bottomrule
\end{tabularx}
\end{table*}

{\footnotesize\setlength{\tabcolsep}{2.5pt}
\begin{longtable}{@{}>{\raggedright\arraybackslash}p{0.27\linewidth}ll>{\raggedright\arraybackslash}p{3.4cm}ccc@{}}
\caption{Included studies by grouping. Modality: SAR, synthetic aperture radar or sonar; Optics, optics and interferometry; Profilometry, profilometry and structured light; Holography, holography and microscopy; MRI, magnetic resonance imaging; Other, other modalities. Merging: what is adjusted per unit (Integer, an integer multiple of $2\pi$; Real, a real-valued offset; None, nothing after the local solves; Other, another variable such as a scale or a tilt), then how the values are obtained (one at a time; all together; during solves, inside the local solves; single integration, after the cuts of all units are joined; vs.\ reference, against an external reference; refused, merging refused by design); n.r., not reported. GT: the result is compared with a phase known independently of the method; another unwrapping algorithm does not count. 1 vs $n$: the same local solver on the same image, run in one piece and on the units. Measured: real data. A blank cell means that the item is absent from the full text.}\label{tab:included}\\
\toprule Study & Year & Modality & Merging & GT & 1 vs $n$ & Measured\\\midrule\endfirsthead
\toprule Study & Year & Modality & Merging & GT & 1 vs $n$ & Measured\\\midrule\endhead
\bottomrule\endfoot
\multicolumn{7}{@{}>{\raggedright\arraybackslash}p{0.68\linewidth}}{G1 \textit{Fixed grid, one integer offset per tile, set sequentially or along a reliability tree} (12)}\\
\citet{towers1991automatic}; also \citet{judge1992holographic,judge1992quantitative,towers1989quasi,quan1991holographic} & 1991 & Holography & Integer, one at a time & \checkmark &  & \checkmark\\
\citet{friedlander1996model} & 1996 & Other & Integer, one at a time & \checkmark &  & \\
\citet{schwarz1999hybrid} & 1999 & Optics & Integer, one at a time & \checkmark &  & \\
\citet{strand1999two} & 1999 & MRI & Integer, one at a time & \checkmark &  & \checkmark\\
\citet{huang2002phase} & 2002 & Optics & Integer, one at a time & \checkmark & \checkmark & \checkmark\\
\citet{lo2002surface}; also \citet{lo2002threedimensional} & 2002 & Profilometry & Integer, one at a time &  & \checkmark & \checkmark\\
\citet{huang2010fast} & 2010 & SAR & Integer, one at a time & \checkmark & \checkmark & \checkmark\\
\citet{antonopoulos2015tilebased}; also \citet{antonopoulos2015experimental} & 2015 & Holography & Integer, one at a time & \checkmark &  & \checkmark\\
\citet{hao2017phase} & 2017 & SAR & Integer, one at a time &  &  & \checkmark\\
\citet{nie2018study} & 2018 & Optics & Integer, n.r. & \checkmark &  & \\
\citet{zhong2019hierarchical} & 2019 & SAR & Integer, one at a time &  & \checkmark & \checkmark\\
\citet{cheng2023novel} & 2023 & MRI & Integer, n.r. & \checkmark &  & \checkmark\\
\addlinespace
\multicolumn{7}{@{}>{\raggedright\arraybackslash}p{0.68\linewidth}}{G2 \textit{Fixed grid, one real-valued offset per tile (seam mean or median, sign and constant), sequential or quality-guided} (7)}\\
\citet{chang1998tapestry} & 1998 & Optics & Real, one at a time &  & \checkmark & \checkmark\\
\citet{dalmaucedeno2008fast} & 2008 & Optics & Other, one at a time &  &  & \checkmark\\
\citet{suo2008data} & 2008 & SAR & Real, n.r. &  &  & \checkmark\\
\citet{wang2013improvement} & 2013 & Holography & Real, one at a time & \checkmark &  & \checkmark\\
\citet{chen2025stitching} & 2025 & Optics & Real, one at a time & \checkmark & \checkmark & \checkmark\\
\citet{moevus2026tilebased} & 2026 & Other & Real, one at a time & \checkmark & \checkmark & \checkmark\\
\citet{onat2026parallel} & 2026 & SAR & Real, one at a time & \checkmark & \checkmark & \checkmark\\
\addlinespace
\multicolumn{7}{@{}>{\raggedright\arraybackslash}p{0.68\linewidth}}{G3 \textit{Adaptive or hierarchical partitions (binary tree, quadtree, k-d tree, content-shifted blocks)} (9)}\\
\citet{arevalilloherraez1996robustsimple}; also \citet{arevalilloherraez1996robust} & 1996 & Optics & Integer, one at a time &  &  & \checkmark\\
\citet{hardie1998robust}; also \citet{blackshire1999advanced} & 1998 & Optics & Integer, one at a time & \checkmark &  & \checkmark\\
\citet{blackshire1999realtime} & 1999 & Optics & Integer, one at a time &  &  & \checkmark\\
\citet{wang1999phaseunwrapping}; also \citet{wang1999efficient} & 1999 & Optics & Integer, n.r. &  &  & \checkmark\\
\citet{baldi2001twodimensional} & 2001 & Optics & Integer, one at a time & \checkmark &  & \checkmark\\
\citet{huang2006phase}; also \citet{huang2006threedimensional} & 2006 & Holography & Integer, one at a time &  &  & \checkmark\\
\citet{wu2007phase} & 2007 & SAR & Real, one at a time & \checkmark & \checkmark & \checkmark\\
\citet{chen2025phase} & 2025 & Optics & Integer, one at a time & \checkmark &  & \checkmark\\
\citet{moevus2026adaptive} & 2026 & Other & Real, one at a time & \checkmark &  & \checkmark\\
\addlinespace
\multicolumn{7}{@{}>{\raggedright\arraybackslash}p{0.68\linewidth}}{G4 \textit{Regions unwrapped by a local solver, then joined by an integer or real-valued offset} (8)}\\
\citet{ching1992two} & 1992 & MRI & Integer, one at a time & \checkmark &  & \\
\citet{hedley1992new} & 1992 & MRI & Integer, one at a time & \checkmark &  & \\
\citet{chen2013improving} & 2013 & Profilometry & Integer, one at a time & \checkmark &  & \checkmark\\
\citet{lu2014fast} & 2014 & Profilometry & Real, one at a time &  & \checkmark & \checkmark\\
\citet{li2015coseismic} & 2015 & SAR & Integer, one at a time & \checkmark & \checkmark & \checkmark\\
\citet{zhang2016relative} & 2016 & Profilometry & Real, n.r. & \checkmark &  & \checkmark\\
\citet{xia2024object} & 2024 & SAR & Integer, one at a time &  &  & \checkmark\\
\citet{ye2024singlefrequency} & 2024 & Profilometry & n.r., one at a time & \checkmark &  & \checkmark\\
\addlinespace
\multicolumn{7}{@{}>{\raggedright\arraybackslash}p{0.68\linewidth}}{G5 \textit{All offsets solved jointly: secondary network, least squares on a graph, QUBO or CQM, discontinuity minimization} (16)}\\
\citet{hung1998phase} & 1998 & Optics & Integer, all together & \checkmark &  & \checkmark\\
\citet{carballo2002hierarchical} & 2002 & SAR & Integer, all together & \checkmark & \checkmark & \checkmark\\
\citet{chen2002phase} & 2002 & SAR & Integer, all together & \checkmark & \checkmark & \checkmark\\
\citet{gao2005phase} & 2005 & SAR & Integer, all together &  &  & \checkmark\\
\citet{zhong2010fast} & 2010 & SAR & Integer, all together &  & \checkmark & \checkmark\\
\citet{zhong2011combined} & 2011 & SAR & Integer, all together &  & \checkmark & \checkmark\\
\citet{zhong2012blocked} & 2012 & SAR & Integer, all together & \checkmark & \checkmark & \checkmark\\
\citet{zhong2013qualityguided} & 2014 & SAR & Integer, all together &  & \checkmark & \checkmark\\
\citet{zhong2018blocked} & 2018 & SAR & Integer, all together &  & \checkmark & \checkmark\\
\citet{zhong2019combined} & 2019 & SAR & Integer, all together &  & \checkmark & \checkmark\\
\citet{kelany2022quantum}; also \citet{kelany2021deep} & 2022 & SAR & Integer, all together & \checkmark & \checkmark & \checkmark\\
\citet{gourdain2024neural} & 2024 & Optics & Real, all together & \checkmark &  & \checkmark\\
\citet{haghighi2024enhanced} & 2024 & SAR & Integer, all together & \checkmark & \checkmark & \\
\citet{wu2024parallel} & 2024 & SAR & n.r., all together & \checkmark & \checkmark & \checkmark\\
\citet{jiang2026wrapcountbased} & 2026 & SAR & Integer, all together & \checkmark & \checkmark & \checkmark\\
\citet{yang2026continuous} & 2026 & SAR & Other, all together & \checkmark &  & \checkmark\\
\addlinespace
\multicolumn{7}{@{}>{\raggedright\arraybackslash}p{0.68\linewidth}}{G6 \textit{Partition imposed by acquisition (subapertures, slices, frames, scenes)} (11)}\\
\citet{ramesh2000phase} & 2000 & Optics & Other, vs.\ reference & \checkmark &  & \checkmark\\
\citet{zhu2006design} & 2006 & Optics & Other, n.r. &  &  & \checkmark\\
\citet{lin2011subaperture} & 2011 & Optics & Other, one at a time & \checkmark &  & \\
\citet{liu2012interventionbased} & 2012 & MRI & Integer, one at a time &  &  & \checkmark\\
\citet{zhao2014study} & 2014 & Optics & Other, one at a time &  & \checkmark & \checkmark\\
\citet{zhou2017topography} & 2017 & Optics & Real, one at a time & \checkmark &  & \checkmark\\
\citet{xie2021phase} & 2021 & Holography & Other, one at a time & \checkmark &  & \checkmark\\
\citet{gao2022experimental} & 2022 & SAR & Other, all together & \checkmark &  & \checkmark\\
\citet{long2022phase} & 2022 & Optics & Real, one at a time & \checkmark &  & \checkmark\\
\citet{zhao2022high}; also \citet{zhao2026phase} & 2022 & Optics & Real, one at a time & \checkmark &  & \checkmark\\
\citet{zhao2026surface} & 2026 & SAR & Other, n.r. & \checkmark &  & \checkmark\\
\addlinespace
\multicolumn{7}{@{}>{\raggedright\arraybackslash}p{0.68\linewidth}}{G7 \textit{Tiling imposed by the fixed input size of a neural network} (5)}\\
\citet{zhao2021phase}; also \citet{zhao2023phase} & 2021 & Optics & Other, one at a time & \checkmark &  & \checkmark\\
\citet{wang2023singlefrequency} & 2023 & Profilometry & Other, vs.\ reference &  &  & \checkmark\\
\citet{liu2024highaccuracy} & 2024 & Profilometry & Other, n.r. & \checkmark &  & \checkmark\\
\citet{wang2024antinoisedesigned} & 2024 & Optics & n.r., n.r. & \checkmark &  & \checkmark\\
\citet{song2026insar} & 2026 & SAR & Other, all together & \checkmark &  & \checkmark\\
\addlinespace
\multicolumn{7}{@{}>{\raggedright\arraybackslash}p{0.68\linewidth}}{G8 \textit{Jump-free regions from the wrapped phase, one integer per region, greedy merging or propagation (PRELUDE lineage and related methods)} (19)}\\
\citet{stetson1997noiseimmune} & 1997 & Optics & Integer, one at a time &  &  & \checkmark\\
\citet{desvignes2000deroulement} & 2000 & MRI & Integer, one at a time &  &  & \checkmark\\
\citet{arevalilloherraez2002robust} & 2002 & Optics & Integer, one at a time &  &  & \checkmark\\
\citet{jenkinson2003fast} & 2003 & MRI & Integer, one at a time & \checkmark &  & \checkmark\\
\citet{siegmann2005robust} & 2005 & Optics & Integer, one at a time & \checkmark &  & \checkmark\\
\citet{wesolkowski2007stochastic} & 2007 & SAR & Integer, one at a time & \checkmark &  & \checkmark\\
\citet{xiong2008phase} & 2008 & SAR & Integer, one at a time &  & \checkmark & \checkmark\\
\citet{liu2014threedimensional} & 2014 & Profilometry & Integer, one at a time &  &  & \checkmark\\
\citet{he2015regiongrowing} & 2015 & MRI & Integer, one at a time & \checkmark &  & \checkmark\\
\citet{liu2016regiongrowing} & 2016 & Optics & Integer, one at a time & \checkmark &  & \checkmark\\
\citet{cheng2017close} & 2018 & MRI & Integer, one at a time & \checkmark &  & \checkmark\\
\citet{cheng2018novel} & 2018 & MRI & Integer, one at a time & \checkmark &  & \checkmark\\
\citet{nie2018fast} & 2018 & Profilometry & Integer, n.r. & \checkmark & \checkmark & \checkmark\\
\citet{karsa2019segue} & 2019 & MRI & Integer, one at a time & \checkmark &  & \checkmark\\
\citet{li2021fast} & 2021 & Optics & Integer, one at a time & \checkmark &  & \checkmark\\
\citet{gao2022phase} & 2022 & SAR & Integer, one at a time &  &  & \checkmark\\
\citet{cheng2022new} & 2023 & MRI & Integer, one at a time & \checkmark &  & \checkmark\\
\citet{cheng2023new} & 2023 & MRI & Integer, one at a time & \checkmark &  & \checkmark\\
\citet{gao2023parallel} & 2023 & SAR & Integer, one at a time &  &  & \checkmark\\
\addlinespace
\multicolumn{7}{@{}>{\raggedright\arraybackslash}p{0.68\linewidth}}{G9 \textit{Jump-free fringe regions, one integer per region (Gierloff type)} (14)}\\
\citet{gierloff1987phase} & 1987 & Optics & Integer, all together & \checkmark &  & \checkmark\\
\citet{charette1996robust} & 1996 & Optics & Integer, one at a time &  &  & \checkmark\\
\citet{joo1997automated} & 1997 & Optics & Integer, one at a time &  &  & \\
\citet{soille2000morphological} & 2000 & Profilometry & Integer, one at a time &  &  & \checkmark\\
\citet{huang2002innovative} & 2002 & Optics & Integer, n.r. &  &  & \checkmark\\
\citet{huang2002quasioneframe} & 2002 & Optics & Integer, n.r. & \checkmark &  & \checkmark\\
\citet{wang2003phase} & 2003 & SAR & Integer, one at a time &  &  & \checkmark\\
\citet{wang2004phase} & 2004 & SAR & Integer, one at a time &  &  & \checkmark\\
\citet{meneses2005phaseunwrapping} & 2005 & Profilometry & Integer, one at a time & \checkmark &  & \checkmark\\
\citet{qiu2006regional} & 2006 & Optics & Integer, one at a time &  &  & \checkmark\\
\citet{huang2008retrieving} & 2008 & Optics & Integer, one at a time & \checkmark &  & \checkmark\\
\citet{huang2010regional} & 2010 & Optics & Integer, one at a time & \checkmark &  & \checkmark\\
\citet{shevkunov2015phase} & 2015 & Holography & Other, one at a time &  &  & \checkmark\\
\citet{wan2026fullfield} & 2026 & Optics & Integer, one at a time & \checkmark &  & \checkmark\\
\addlinespace
\multicolumn{7}{@{}>{\raggedright\arraybackslash}p{0.68\linewidth}}{G10 \textit{Integer region labels chosen by an energy on the region graph (graph cut, Markov random field)} (4)}\\
\citet{xie2010patchbased} & 2010 & MRI & Integer, all together & \checkmark &  & \checkmark\\
\citet{wang2021phase} & 2021 & Other & Integer, n.r. & \checkmark &  & \checkmark\\
\citet{zhou2022canet} & 2022 & SAR & Integer, all together & \checkmark & \checkmark & \checkmark\\
\citet{yang2025phaseconnectedregionbased} & 2025 & Profilometry & Integer, all together & \checkmark &  & \checkmark\\
\addlinespace
\multicolumn{7}{@{}>{\raggedright\arraybackslash}p{0.68\linewidth}}{G11 \textit{Blocks with imposed values, or a coarse global solution computed before partitioning, mostly without offset estimation} (5)}\\
\citet{costantini1998novel} & 1998 & SAR & None, during solves & \checkmark &  & \checkmark\\
\citet{pepe2011new} & 2011 & SAR & None, during solves & \checkmark & \checkmark & \checkmark\\
\citet{zhang2011phase} & 2011 & SAR & None, during solves & \checkmark & \checkmark & \checkmark\\
\citet{costantini2012multiscale} & 2012 & SAR & Integer, n.r. &  &  & \checkmark\\
\citet{zhang2018repeatpass} & 2018 & SAR & None, during solves & \checkmark &  & \checkmark\\
\addlinespace
\multicolumn{7}{@{}>{\raggedright\arraybackslash}p{0.68\linewidth}}{G12 \textit{Reliability regions unwrapped in order, each anchored by Dirichlet conditions or constraints from the area already unwrapped} (9)}\\
\citet{fornaro1999twodimensional} & 1999 & SAR & None, during solves & \checkmark & \checkmark & \checkmark\\
\citet{fornaro2001generation} & 2001 & SAR & n.r., vs.\ reference & \checkmark &  & \checkmark\\
\citet{wu2004insar} & 2004 & SAR & None, during solves & \checkmark &  & \\
\citet{zhang2009new} & 2009 & SAR & None, during solves & \checkmark &  & \\
\citet{xu2015phase} & 2015 & SAR & None, during solves &  & \checkmark & \checkmark\\
\citet{zhong2016combined} & 2016 & SAR & None, during solves &  &  & \checkmark\\
\citet{zhang2018hybrid} & 2018 & SAR & None, one at a time &  &  & \checkmark\\
\citet{zhong2018phase} & 2018 & SAR & None, during solves & \checkmark &  & \checkmark\\
\citet{bao2022phase} & 2022 & Profilometry & None, one at a time & \checkmark &  & \checkmark\\
\addlinespace
\multicolumn{7}{@{}>{\raggedright\arraybackslash}p{0.68\linewidth}}{G13 \textit{Coordination during solving or exact resolution: no merging afterwards (synchronized strips, warm start, dual decomposition)} (3)}\\
\citet{huang2015parallel} & 2015 & SAR & None, during solves & \checkmark & \checkmark & \checkmark\\
\citet{dong2016phase} & 2017 & MRI & None, during solves &  & \checkmark & \checkmark\\
\citet{barabadi2019dualstage} & 2019 & SAR & None, during solves &  & \checkmark & \checkmark\\
\addlinespace
\multicolumn{7}{@{}>{\raggedright\arraybackslash}p{0.68\linewidth}}{G14 \textit{Partitions that cut no residue cluster: local branch cuts joined, then a single integration} (6)}\\
\citet{yu2011residues} & 2011 & SAR & None, during solves & \checkmark & \checkmark & \checkmark\\
\citet{yu2013fast} & 2013 & SAR & None, single integration & \checkmark & \checkmark & \checkmark\\
\citet{yu2017largescale} & 2017 & SAR & None, single integration &  & \checkmark & \checkmark\\
\citet{yu2019largescale} & 2019 & SAR & None, single integration & \checkmark & \checkmark & \checkmark\\
\citet{lan2023clusteranalysis} & 2023 & SAR & None, single integration & \checkmark &  & \checkmark\\
\citet{lan2025elaborate} & 2025 & SAR & n.r., n.r. & \checkmark &  & \checkmark\\
\addlinespace
\multicolumn{7}{@{}>{\raggedright\arraybackslash}p{0.68\linewidth}}{G15 \textit{Units tied to an external absolute reference, without a shared seam} (8)}\\
\citet{liu2007synergistic} & 2007 & SAR & Real, vs.\ reference &  &  & \checkmark\\
\citet{cui2011reliabilityguided} & 2011 & Profilometry & Other, vs.\ reference & \checkmark &  & \checkmark\\
\citet{liu2015clusteranalysisbased} & 2015 & SAR & Integer, vs.\ reference &  &  & \checkmark\\
\citet{wang2016using} & 2016 & SAR & Integer, all together & \checkmark &  & \checkmark\\
\citet{libert2017splitband} & 2017 & SAR & Integer, vs.\ reference &  & \checkmark & \checkmark\\
\citet{yi2025novel} & 2025 & Profilometry & Integer, vs.\ reference & \checkmark & \checkmark & \checkmark\\
\citet{yuan2025waveclusterbased} & 2025 & SAR & Integer, vs.\ reference & \checkmark &  & \checkmark\\
\citet{zhao2026wetland} & 2026 & SAR & Real, vs.\ reference &  & \checkmark & \checkmark\\
\addlinespace
\multicolumn{7}{@{}>{\raggedright\arraybackslash}p{0.68\linewidth}}{G16 \textit{Merging deliberately refused because the discontinuity is physical} (1)}\\
\citet{song2024toward}; also \citet{song2026monitoring} & 2024 & SAR & None, refused &  &  & \checkmark\\
\addlinespace
\multicolumn{7}{@{}>{\raggedright\arraybackslash}p{0.68\linewidth}}{G17 \textit{Partition described, merging not described or deferred to another source} (12)}\\
\citet{stephenson1994data} & 1994 & Optics & n.r., n.r. &  & \checkmark & \checkmark\\
\citet{saiprasad2004towards} & 2004 & Optics & n.r., n.r. & \checkmark &  & \checkmark\\
\citet{sawaf2006finding} & 2006 & Optics & n.r., n.r. &  &  & \\
\citet{huang2007phase} & 2007 & Optics & n.r., one at a time &  &  & \checkmark\\
\citet{madhu2007new} & 2007 & Optics & n.r., one at a time &  &  & \checkmark\\
\citet{dudczyk2014optimizing} & 2014 & SAR & n.r., n.r. &  &  & \checkmark\\
\citet{huang2015groundbased} & 2015 & SAR & n.r., n.r. &  &  & \checkmark\\
\citet{kulkarni2015threedimensional}; also \citet{kulkarni2016spatial} & 2015 & Holography & n.r., n.r. & \checkmark &  & \checkmark\\
\citet{zhao2015snakeassisted}; also \citet{zhao2016windowed} & 2015 & Profilometry & n.r., n.r. & \checkmark & \checkmark & \checkmark\\
\citet{gao2017twodimensional} & 2017 & SAR & n.r., n.r. & \checkmark &  & \checkmark\\
\citet{khodadad2025fast} & 2025 & Holography & n.r., n.r. & \checkmark &  & \checkmark\\
\citet{zhang2025factor} & 2025 & SAR & n.r., n.r. & \checkmark &  & \checkmark\\
\addlinespace
\end{longtable}}

\paragraph{Notes on some codes.}
Gierloff is coded ``all together'' because the method infers the offsets from the relations between all regions before it assigns any of them, although the algorithm is not specified~\citep{gierloff1987phase}. The later studies of its grouping fix the regions one at a time, so Figure~\ref{fig:hierarchy} places the grouping with the methods that choose region by region. Song et al. (2026) describe no offset between tiles: their maps are averaged rather than shifted, which is coded ``Other''~\citep{song2026insar}. Wang et al. (2023) also describe no offset, and align the merged map row by row on a separately measured reference plane~\citep{wang2023singlefrequency}. Both stay with the network tilings of class~I, because their units are the tiles of a network input. The fixed hierarchies of grouping G3 count as fixed geometric tiles in Table~\ref{tab:characteristics}, because their trees ignore the phase.

\section{Reports excluded at full text}\label{app:excluded}
Table~\ref{tab:excluded} lists the \nExcludedFullText reports that were read in full and excluded, by year, each with its reason (Table~\ref{tab:exclusions}).
{\footnotesize\setlength{\tabcolsep}{3pt}
\begin{longtable}{@{}l>{\raggedright\arraybackslash}p{0.62\linewidth}>{\raggedright\arraybackslash}p{0.27\linewidth}@{}}
\caption{Reports excluded at full text, with the reason for each exclusion. A title in Chinese is replaced by a note; the identifier then locates the report.}\label{tab:excluded}\\
\toprule
Year & Report & Reason\\
\midrule
\endfirsthead
\toprule
Year & Report & Reason\\
\midrule
\endhead
\bottomrule
\endlastfoot
1988 & A windowing technique for the automated analysis of holo-interferograms. \href{https://doi.org/10.1016/0143-8166(88)90048-6}{\nolinkurl{10.1016/0143-8166(88)90048-6}} & No partition fixed before the local solves\\
1991 & Investigation of the cellular automata method for phase unwrapping and its implementation on an array processor. \href{https://doi.org/10.1016/0143-8166(91)90036-s}{\nolinkurl{10.1016/0143-8166(91)90036-s}} & No partition fixed before the local solves\\
1992 & Practical phase unwrapping of holographic interferograms. \href{https://doi.org/10.1117/12.135306}{\nolinkurl{10.1117/12.135306}} & No partition fixed before the local solves\\
1994 & A hybrid fringe analysis technique for the elimination of random noise in interferometric wrapped phase maps. \href{https://doi.org/10.1016/0030-4018(94)90455-3}{\nolinkurl{10.1016/0030-4018(94)90455-3}} & No phase unwrapping, or partition without unwrapping\\
1995 & Noise-robust phase-unwrapping method in radar interferometry. \href{https://doi.org/10.1117/12.227140}{\nolinkurl{10.1117/12.227140}} & No partition fixed before the local solves\\
1995 & Unweighted least squares phase unwrapping by means of multigrid techniques. \href{https://doi.org/10.1117/12.227138}{\nolinkurl{10.1117/12.227138}} & No partition fixed before the local solves\\
1997 & A noise-immune method of phase unwrapping in speckle interferometry. \href{https://doi.org/10.1016/0143-8166(95)00110-7}{\nolinkurl{10.1016/0143-8166(95)00110-7}} & No partition fixed before the local solves\\
1997 & Phase unwrapping algorithm based on reliability and edge detection. \href{https://doi.org/10.1117/1.601194}{\nolinkurl{10.1117/1.601194}} & No partition fixed before the local solves\\
1997 & The Developing of a Wide-Area Interferometric Processor. & No partition fixed before the local solves\\
1998 & Phase unwrapping through a branch-cut based recursive patching method. \href{https://doi.org/10.1109/igarss.1998.691476}{\nolinkurl{10.1109/igarss.1998.691476}} & Units created during unwrapping\\
1998 & Two-Dimensional Phase Unwrapping by Direct Elimination of Rotational Vector Fields from Phase Gradients Obtained by Heterodyne Techniques. \href{https://doi.org/10.1007/s10043-998-0374-4}{\nolinkurl{10.1007/s10043-998-0374-4}} & No partition fixed before the local solves\\
1999 & Phase unwrapping through a branch-cut-based cut-bridging and window-patching method. \href{https://doi.org/10.1364/ao.38.000805}{\nolinkurl{10.1364/ao.38.000805}} & Units created during unwrapping\\
1999 & Piecewise smooth phase reconstruction. \href{https://doi.org/10.1364/ol.24.001802}{\nolinkurl{10.1364/ol.24.001802}} & Units created during unwrapping\\
2001 & Weighted multi-resolution phase-unwrapping method. & No partition fixed before the local solves\\
2002 & New Robust 3-D Phase Unwrapping Algorithms: Application to Magnetic Field Mapping and Undistorting Echoplanar Images. \href{https://doi.org/10.1006/nimg.2002.1092}{\nolinkurl{10.1006/nimg.2002.1092}} & Units created during unwrapping\\
2002 & Noise immune unwrapping based on phase statistics and self-calibration. \href{https://doi.org/10.1016/s0143-8166(02)00023-4}{\nolinkurl{10.1016/s0143-8166(02)00023-4}} & No partition fixed before the local solves\\
2002 & Phase unwrapping based on a parallel noise-immune algorithm. \href{https://doi.org/10.1016/s0030-3992(02)00042-7}{\nolinkurl{10.1016/s0030-3992(02)00042-7}} & No partition fixed before the local solves\\
2003 & A three-dimensional phase unwrapping algorithm for processing of multitemporal SAR interferometric measurements. \href{https://doi.org/10.1109/igarss.2002.1026239}{\nolinkurl{10.1109/igarss.2002.1026239}} & No partition fixed before the local solves\\
2003 & Multi-resolution phase model reconstruction for phase unwrapping. \href{https://doi.org/10.1109/igarss.1999.771521}{\nolinkurl{10.1109/igarss.1999.771521}} & No partition fixed before the local solves\\
2003 & Phase unwrapping by region growing. \href{https://doi.org/10.1364/ao.42.002498}{\nolinkurl{10.1364/ao.42.002498}} & Units created during unwrapping\\
2004 & Superresolution of Ultrasound Images Using the First and Second Harmonic Signal. \href{https://doi.org/10.1109/tuffc.2004.1295391}{\nolinkurl{10.1109/tuffc.2004.1295391}} & Units never merged into one phase field\\
2005 & SAR Interferometry: A 2-D Phase Unwrapping Technique Based On Phase And Absolute Values Informations. \href{https://doi.org/10.1109/igarss.1990.688929}{\nolinkurl{10.1109/igarss.1990.688929}} & No partition fixed before the local solves\\
2006 & Modifled nearest neighbor phase unwrapping algorilthm. \href{https://doi.org/10.1007/bf03033668}{\nolinkurl{10.1007/bf03033668}} & No partition fixed before the local solves\\
2006 & Multilevel quality-guided phase unwrapping algorithm for real-time three-dimensional shape reconstruction. \href{https://doi.org/10.1364/ao.46.000050}{\nolinkurl{10.1364/ao.46.000050}} & No partition fixed before the local solves\\
2006 & Two-dimensional windowed Fourier transform for fringe pattern analysis: Principles, applications and implementations. \href{https://doi.org/10.1016/j.optlaseng.2005.10.012}{\nolinkurl{10.1016/j.optlaseng.2005.10.012}} & No partition fixed before the local solves\\
2007 & AN N -DIMENSIONAL RECURRENT NOISY PHASE UNWRAPPING ALGORITHM. \href{https://doi.org/10.2316/journal.202.2007.4.202-2232}{\nolinkurl{10.2316/journal.202.2007.4.202-2232}} & No partition fixed before the local solves\\
2007 & Phase unwrapping in three dimensions with application to InSAR time series. \href{https://doi.org/10.1364/josaa.24.002737}{\nolinkurl{10.1364/josaa.24.002737}} & Non-spatial decomposition\\
2008 & Simple and accurate unwrapping phase of MR data. \href{https://doi.org/10.1016/j.measurement.2008.12.003}{\nolinkurl{10.1016/j.measurement.2008.12.003}} & No partition fixed before the local solves\\
2009 & Piecewise polynomial phase approximation approach for the analysis of reconstructed interference fields in digital holographic interferometry. \href{https://doi.org/10.1088/1464-4258/11/6/065405}{\nolinkurl{10.1088/1464-4258/11/6/065405}} & One-dimensional unwrapping\\
2009 & Windowed high-order ambiguity function method for fringe analysis. \href{https://doi.org/10.1063/1.3186646}{\nolinkurl{10.1063/1.3186646}} & One-dimensional unwrapping\\
2010 & A Cluster-Analysis-Based Efficient Multibaseline Phase-Unwrapping Algorithm. \href{https://doi.org/10.1109/tgrs.2010.2055569}{\nolinkurl{10.1109/tgrs.2010.2055569}} & Non-spatial decomposition\\
2010 & A New Strategy to Estimate Local Fringe Frequencies for InSAR Phase Noise Reduction. \href{https://doi.org/10.1109/lgrs.2010.2047935}{\nolinkurl{10.1109/lgrs.2010.2047935}} & Units never merged into one phase field\\
2010 & Fourier Transform Interferometry for 3D Mapping of Rough and Discontinuous Surfaces. & Units never merged into one phase field\\
2010 & Method of phase unwrapping-free DEM reconstruction of InSAS. \href{https://doi.org/10.1109/iasp.2010.5476163}{\nolinkurl{10.1109/iasp.2010.5476163}} & No partition fixed before the local solves\\
2010 & Multi-patch and centre-expansion algorithms for phase unwrapping of large InSAR images with dense residues. \href{https://doi.org/10.1080/01431160903095452}{\nolinkurl{10.1080/01431160903095452}} & No partition fixed before the local solves\\
2011 & A fast, accurate phase unwrapping method for wavelet-transform profilometry. \href{https://doi.org/10.1016/j.optcom.2011.10.012}{\nolinkurl{10.1016/j.optcom.2011.10.012}} & No partition fixed before the local solves\\
2011 & Quality-guided phase unwrapping algorithm based on reliability evaluation. \href{https://doi.org/10.1364/ao.50.005446}{\nolinkurl{10.1364/ao.50.005446}} & No partition fixed before the local solves\\
2012 & An application of the principle of minimal frustration to phase unwrapping. \href{https://doi.org/10.1016/j.dsp.2012.12.002}{\nolinkurl{10.1016/j.dsp.2012.12.002}} & No partition fixed before the local solves\\
2012 & Consistent Stereo-Assisted Absolute Phase Unwrapping Methods for Structured Light Systems. \href{https://doi.org/10.1109/jstsp.2012.2195157}{\nolinkurl{10.1109/jstsp.2012.2195157}} & Units created during unwrapping\\
2012 & Fast phase unwrapping algorithm based on partition of dense residual regions. \href{https://doi.org/10.3873/j.issn.1000-1328.2012.02.007}{\nolinkurl{10.3873/j.issn.1000-1328.2012.02.007}} & No partition fixed before the local solves\\
2012 & GIS-supported iterative filtering approach for building height estimation from InSAR data. & Application of an existing unwrapper\\
2012 & Phase unwrapping using discrete Particle Swarm Optimization. & No partition fixed before the local solves\\
2012 & Three-dimensional shape measurement of a highly reflected, specular surface with structured light method. \href{https://doi.org/10.1364/ao.51.007724}{\nolinkurl{10.1364/ao.51.007724}} & No partition fixed before the local solves\\
2012 & Windowed phase unwrapping using a first-order dynamic system following iso-phase contours. \href{https://doi.org/10.1364/ao.51.007549}{\nolinkurl{10.1364/ao.51.007549}} & No partition fixed before the local solves\\
2013 & Parallel phase unwrapping in 3D shape measurement using digital fringe projection technique. \href{https://doi.org/10.1109/icip.2013.6738657}{\nolinkurl{10.1109/icip.2013.6738657}} & No partition fixed before the local solves\\
2013 & Real-time Unwrapped Phase Generating Algorithm based on Airborne Dual-antenna InSAR System. \href{https://doi.org/10.4304/jsw.8.11.2956-2961}{\nolinkurl{10.4304/jsw.8.11.2956-2961}} & No partition fixed before the local solves\\
2013 & Three-Dimensional Phase Unwrapping for Satellite Radar Interferometry, I: DEM Generation. \href{https://doi.org/10.1109/tgrs.2013.2247043}{\nolinkurl{10.1109/tgrs.2013.2247043}} & No partition fixed before the local solves\\
2014 & Algebraic phase unwrapping over collection of triangles based on two-dimensional spline smoothing. \href{https://doi.org/10.1109/icassp.2014.6854546}{\nolinkurl{10.1109/icassp.2014.6854546}} & No partition fixed before the local solves\\
2014 & An Approach to Persistent Scatterer Interferometry. \href{https://doi.org/10.3390/rs6076662}{\nolinkurl{10.3390/rs6076662}} & Non-spatial decomposition\\
2014 & High-resolution deformation measurement using "Persistent Scatterer Interferometry". \href{https://doi.org/10.5821/dissertation-2117-95500}{\nolinkurl{10.5821/dissertation-2117-95500}} & No partition fixed before the local solves\\
2014 & Real-time 4D phase unwrapping applied to magnetic resonance elastography. \href{https://doi.org/10.1002/mrm.25332}{\nolinkurl{10.1002/mrm.25332}} & No partition fixed before the local solves\\
2014 & Region-referenced phase unwrapping architecture for digital holographic microscopy. \href{https://doi.org/10.1364/ao.54.000a67}{\nolinkurl{10.1364/ao.54.000a67}} & No partition fixed before the local solves\\
2014 & Robust phase unwrapping for MR temperature imaging using a magnitude-sorted list, multi-clustering algorithm. \href{https://doi.org/10.1002/mrm.25279}{\nolinkurl{10.1002/mrm.25279}} & Units created during unwrapping\\
2015 & A Phase Unwrapping Algorithm for Interferometric Phase Images. \href{https://doi.org/10.1007/978-3-319-26123-2_14}{\nolinkurl{10.1007/978-3-319-26123-2_14}} & No partition fixed before the local solves\\
2015 & A novel algorithm based on histogram processing of reliability for two-dimensional phase unwrapping. \href{https://doi.org/10.1016/j.ijleo.2015.04.070}{\nolinkurl{10.1016/j.ijleo.2015.04.070}} & Units created during unwrapping\\
2015 & A virtual resampling technique for algebraic two-dimensional phase unwrapping. \href{https://doi.org/10.1109/icassp.2015.7178696}{\nolinkurl{10.1109/icassp.2015.7178696}} & No partition fixed before the local solves\\
2015 & Combination of temporal phase unwrapping and long-wave infrared digital holographic interferometry for metrology of mosaic detector under space simulated conditions. \href{https://doi.org/10.1117/12.2195896}{\nolinkurl{10.1117/12.2195896}} & One-dimensional unwrapping\\
2015 & Hybrid random walk-linear discriminant analysis method for unwrapping quantitative phase microscopy images of biological samples. \href{https://doi.org/10.1117/1.jbo.20.11.111211}{\nolinkurl{10.1117/1.jbo.20.11.111211}} & No partition fixed before the local solves\\
2015 & One shot profilometry using phase partitions. \href{https://doi.org/10.1016/j.optlaseng.2014.12.016}{\nolinkurl{10.1016/j.optlaseng.2014.12.016}} & Non-spatial decomposition\\
2015 & Phase unwrapping algorithm based on double-frequency fringe projection and fringe background. \href{https://doi.org/10.1117/12.2075884}{\nolinkurl{10.1117/12.2075884}} & No partition fixed before the local solves\\
2015 & Weighted robust Basis Function for phase unwrapping. \href{https://doi.org/10.1016/j.optlaseng.2015.01.010}{\nolinkurl{10.1016/j.optlaseng.2015.01.010}} & No partition fixed before the local solves\\
2016 & A novel fast phase unwrapping method for large interferometric datasets. \href{https://doi.org/10.1109/igarss.2016.7730684}{\nolinkurl{10.1109/igarss.2016.7730684}} & No partition fixed before the local solves\\
2016 & Accelerating phase unwrapping based on integer linear programming by processing of subgraphs. \href{https://doi.org/10.1109/igarss.2016.7730683}{\nolinkurl{10.1109/igarss.2016.7730683}} & Non-spatial decomposition\\
2016 & Pixel-by-pixel absolute phase retrieval using three phase-shifted fringe patterns without markers. \href{https://doi.org/10.1016/j.optlaseng.2016.12.002}{\nolinkurl{10.1016/j.optlaseng.2016.12.002}} & Non-spatial decomposition\\
2016 & Simultaneous estimation of unwrapped phase and phase derivative from a closed fringe pattern. \href{https://doi.org/10.1016/j.optlaseng.2016.03.016}{\nolinkurl{10.1016/j.optlaseng.2016.03.016}} & No partition fixed before the local solves\\
2017 & A Refined Cluster-Analysis-Based Multibaseline Phase-Unwrapping Algorithm. \href{https://doi.org/10.1109/lgrs.2017.2723050}{\nolinkurl{10.1109/lgrs.2017.2723050}} & Non-spatial decomposition\\
2017 & A convex hull algorithm based fast large-scale two-dimensional phase unwrapping method. \href{https://doi.org/10.1109/igarss.2017.8127834}{\nolinkurl{10.1109/igarss.2017.8127834}} & No partition fixed before the local solves\\
2017 & Geographical Situation Monitoring Applications Based on MiniSAR. \href{https://doi.org/10.1007/978-981-10-3966-9_40}{\nolinkurl{10.1007/978-981-10-3966-9_40}} & Application of an existing unwrapper\\
2017 & Phase unwrapping based on radial basis function with adaptive parameters. \href{https://doi.org/10.16136/j.joel.2017.03.0266}{\nolinkurl{10.16136/j.joel.2017.03.0266}} & No partition fixed before the local solves\\
2017 & Three-dimensional Fourier Fringe Analysis and phase unwrapping. \href{https://doi.org/10.24377/ljmu.t.00005846}{\nolinkurl{10.24377/ljmu.t.00005846}} & Units created during unwrapping\\
2017 & Two-dimensional phase unwrapping. \href{https://doi.org/10.24377/ljmu.t.00005814}{\nolinkurl{10.24377/ljmu.t.00005814}} & No partition fixed before the local solves\\
2018 & A Phase Unwrapping Algorithm Based on Branch-cut and B-Spline Fitting in InSAR. \href{https://doi.org/10.1109/icspcc.2018.8567735}{\nolinkurl{10.1109/icspcc.2018.8567735}} & No partition fixed before the local solves\\
2018 & PURE: Scalable Phase Unwrapping with Spatial Redundant Arcs. \href{https://doi.org/10.48550/arxiv.1805.00321}{\nolinkurl{10.48550/arxiv.1805.00321}} & Non-spatial decomposition\\
2018 & Stereo vision and fourier transform profilometry for 3D measurement. \href{https://doi.org/10.1117/12.2319564}{\nolinkurl{10.1117/12.2319564}} & No phase unwrapping, or partition without unwrapping\\
2019 & A novel algorithm based on sub-fringe integration method for direct two-dimensional unwrapping phase reconstruction from the intensity of one-shot two-beam interference fringes. \href{https://doi.org/10.1007/s00340-019-7330-9}{\nolinkurl{10.1007/s00340-019-7330-9}} & No partition fixed before the local solves\\
2019 & A parallel path-following phase unwrapping algorithm based on a top-down breadth-first search approach. \href{https://doi.org/10.1016/j.optlaseng.2019.105827}{\nolinkurl{10.1016/j.optlaseng.2019.105827}} & No partition fixed before the local solves\\
2019 & New particle filtering phase unwrapping algorithm for wrapped fringe patterns. \href{https://doi.org/10.1016/j.optlaseng.2019.01.001}{\nolinkurl{10.1016/j.optlaseng.2019.01.001}} & No partition fixed before the local solves\\
2019 & Small baseline InSAR time series analysis: Unwrapping error correction and noise reduction. \href{https://doi.org/10.31223/osf.io/9sz6m}{\nolinkurl{10.31223/osf.io/9sz6m}} & Units created during unwrapping\\
2020 & A Closed-Form Robust Cluster-Analysis-Based Multibaseline InSAR Phase Unwrapping and Filtering Algorithm With Optimal Baseline Combination Analysis. \href{https://doi.org/10.1109/tgrs.2019.2962001}{\nolinkurl{10.1109/tgrs.2019.2962001}} & Non-spatial decomposition\\
2020 & An Adaptive Statistical Multi-grid DInSAR Technique for Studying Multi-scale Earth Surface Deformation Phenomena. \href{https://doi.org/10.1109/igarss39084.2020.9324535}{\nolinkurl{10.1109/igarss39084.2020.9324535}} & Resolution hierarchy without side-by-side units\\
2020 & Improved fast phase unwrapping algorithm based on parallel acceleration. \href{https://doi.org/10.12086/oee.2020.200111}{\nolinkurl{10.12086/oee.2020.200111}} & Units created during unwrapping\\
2021 & A Parallel Implementation of The Rotational Compensator Phase Unwrapping Algorithm. \href{https://doi.org/10.46719/npsc20212912}{\nolinkurl{10.46719/npsc20212912}} & No partition fixed before the local solves\\
2021 & A phase unwrapping method suitable for high frequency fringe based on edge feature. \href{https://doi.org/10.1016/j.measurement.2021.109938}{\nolinkurl{10.1016/j.measurement.2021.109938}} & No partition fixed before the local solves\\
2021 & An Improved Phase Unwrapping Method Based on Hierarchical Networking and Constrained Adjustment. \href{https://doi.org/10.3390/rs13214193}{\nolinkurl{10.3390/rs13214193}} & No partition fixed before the local solves\\
2021 & Deep Learning for the Detection and Phase Unwrapping of Mining-Induced Deformation in Large-Scale Interferograms. \href{https://doi.org/10.1109/tgrs.2021.3121907}{\nolinkurl{10.1109/tgrs.2021.3121907}} & Units never merged into one phase field\\
2021 & Enhanced geometric constraint-based phase unwrapping algorithm in binocular stereo vision fringe projection system. \href{https://doi.org/10.1784/insi.2021.63.9.540}{\nolinkurl{10.1784/insi.2021.63.9.540}} & Units created during unwrapping\\
2021 & InSAR Phase Unwrapping Error Correction for Rapid Repeat Measurements of Water Level Change in Wetlands. \href{https://doi.org/10.1109/tgrs.2021.3108751}{\nolinkurl{10.1109/tgrs.2021.3108751}} & Units created during unwrapping\\
2021 & Phase unmixing of TerraSAR-X staring spotlight interferograms in building scale for PS height and deformation. \href{https://doi.org/10.1016/j.isprsjprs.2021.08.007}{\nolinkurl{10.1016/j.isprsjprs.2021.08.007}} & Units never merged into one phase field\\
2022 & A Multicomponent Temporal Coherence Model for 3-D Phase Unwrapping in Time-Series InSAR of Seasonal Deformation Areas. \href{https://doi.org/10.3390/rs14051080}{\nolinkurl{10.3390/rs14051080}} & Non-spatial decomposition\\
2022 & BM3D Denoising for a Cluster-Analysis-Based Multibaseline InSAR Phase-Unwrapping Method. \href{https://doi.org/10.3390/rs14081836}{\nolinkurl{10.3390/rs14081836}} & Non-spatial decomposition\\
2022 & Error analysis and 3D reconstruction using airborne array InSAR images. \href{https://doi.org/10.1016/j.isprsjprs.2022.06.005}{\nolinkurl{10.1016/j.isprsjprs.2022.06.005}} & No partition fixed before the local solves\\
2022 & Phase unwrapping algorithm for images with local high-density noise. \href{https://doi.org/10.1364/ao.472005}{\nolinkurl{10.1364/ao.472005}} & No partition fixed before the local solves\\
2023 & InSAR Phase Unwrapping Algorithm Based on Deep GAN. \href{https://doi.org/10.1155/2023/9230780}{\nolinkurl{10.1155/2023/9230780}} & Retracted by the publisher\\
2023 & Inter-partition phase unwrapping method based on Gray code. \href{https://doi.org/10.5768/jao202344.0102003}{\nolinkurl{10.5768/jao202344.0102003}} & Non-spatial decomposition\\
2023 & Ultra-robust spatial point-to-point phase unwrapping algorithm for severe interference signal in 3D sensing. \href{https://doi.org/10.1016/j.sna.2023.114156}{\nolinkurl{10.1016/j.sna.2023.114156}} & No partition fixed before the local solves\\
2024 & 3D automatic detection and correction for phase unwrapping errors in time series SAR interferometry. \href{https://doi.org/10.1016/j.isprsjprs.2024.12.013}{\nolinkurl{10.1016/j.isprsjprs.2024.12.013}} & No partition fixed before the local solves\\
2024 & A Novel Knowledge-Learning Coupling Method for InSAR Phase Unwrapping of Large Surface Displacements in Coal Mining Areas. \href{https://doi.org/10.1109/tgrs.2024.3492505}{\nolinkurl{10.1109/tgrs.2024.3492505}} & No partition fixed before the local solves\\
2024 & [Title in Chinese]. \href{https://doi.org/10.11947/j.agcs.2024.20220617}{\nolinkurl{10.11947/j.agcs.2024.20220617}} & Non-spatial decomposition\\
2024 & An Algorithm for Solving the Problem of Phase Unwrapping in Remote Sensing Radars and Its Implementation on Multicore Processors. \href{https://doi.org/10.3390/math12050727}{\nolinkurl{10.3390/math12050727}} & No partition fixed before the local solves\\
2024 & Block Clustering Method for Large Scale InSAR Interferometric Phase Diagram based on GSOM. \href{https://doi.org/10.1109/eit63098.2024.10762717}{\nolinkurl{10.1109/eit63098.2024.10762717}} & No phase unwrapping, or partition without unwrapping\\
2024 & [Title in Chinese]. \href{https://doi.org/10.14188/j.2095-6045.20230429}{\nolinkurl{10.14188/j.2095-6045.20230429}} & Application of an existing unwrapper\\
2024 & Estimating unwrapped phase through phase gradients. \href{https://doi.org/10.1016/j.ijleo.2024.171901}{\nolinkurl{10.1016/j.ijleo.2024.171901}} & No partition fixed before the local solves\\
2024 & Parallel optimization design of phase unwrapping algorithm. \href{https://doi.org/10.1117/12.3035194}{\nolinkurl{10.1117/12.3035194}} & No partition fixed before the local solves\\
2024 & Phase unwrapping algorithm based on phase diversity wavefront reconstruction and virtual Hartmann--Shack technology. \href{https://doi.org/10.1364/ol.515821}{\nolinkurl{10.1364/ol.515821}} & No partition fixed before the local solves\\
2024 & Phase unwrapping via fully exploiting global and local spatial dependencies. \href{https://doi.org/10.1016/j.optlastec.2024.111872}{\nolinkurl{10.1016/j.optlastec.2024.111872}} & No partition fixed before the local solves\\
2024 & Phase unwrapping via hierarchical and balanced residue partitioning. \href{https://doi.org/10.1007/s11760-023-02958-5}{\nolinkurl{10.1007/s11760-023-02958-5}} & No partition fixed before the local solves\\
2024 & Robust phase unwrapping algorithm for gear interferometry based on mutual information and adaptive spin-inpainting approaches. \href{https://doi.org/10.1016/j.optcom.2024.130748}{\nolinkurl{10.1016/j.optcom.2024.130748}} & No partition fixed before the local solves\\
2024 & Spatial phase unwrapping algorithm based on multi-anchors bidirectional location and suppression. \href{https://doi.org/10.1117/1.oe.63.1.014101}{\nolinkurl{10.1117/1.oe.63.1.014101}} & No partition fixed before the local solves\\
2025 & An SBAS-InSAR Phase Unwrapping Method Integrating ICU and SNAPHU: A Case Study of Dalian City. \href{https://doi.org/10.1109/icivc66358.2025.11200201}{\nolinkurl{10.1109/icivc66358.2025.11200201}} & No partition fixed before the local solves\\
2025 & DIP-UP: Deep Image Prior for Unwrapping Phase. \href{https://doi.org/10.3390/info16070592}{\nolinkurl{10.3390/info16070592}} & Patches used only to train a network\\
2025 & Hybrid CNN-ViT for Phase Unwrapping: A Segmentation-Based Approach. \href{https://doi.org/10.1109/ingarss67683.2025.11584000}{\nolinkurl{10.1109/ingarss67683.2025.11584000}} & No partition fixed before the local solves\\
2025 & Phase unwrapping following a multi-seed short-path constructed by undirected graphs. \href{https://doi.org/10.1364/ol.570484}{\nolinkurl{10.1364/ol.570484}} & Units created during unwrapping\\
2025 & Phase unwrapping for projection moiré based on quaternary equal-width code. \href{https://doi.org/10.1016/j.optlaseng.2025.108844}{\nolinkurl{10.1016/j.optlaseng.2025.108844}} & Non-spatial decomposition\\
2025 & Positioning and Mitigating Mining-Induced Phase Gradients for InSAR Phase Unwrapping. \href{https://doi.org/10.1109/jstars.2025.3560139}{\nolinkurl{10.1109/jstars.2025.3560139}} & No partition fixed before the local solves\\
2025 & Robust cutting plane pure integer programming phase unwrapping algorithm considering fringe frequency for dual-baseline InSAR. \href{https://doi.org/10.1016/j.isprsjprs.2025.05.017}{\nolinkurl{10.1016/j.isprsjprs.2025.05.017}} & Non-spatial decomposition\\
2025 & UMSPU: Universal multi-size phase unwrapping network via mutual self-distillation and adaptive boosting ensemble segmenter. \href{https://doi.org/10.1016/j.optlastec.2025.113916}{\nolinkurl{10.1016/j.optlastec.2025.113916}} & No partition fixed before the local solves\\
2026 & An Improved Closed-Form Multi-Baseline Phase Unwrapping Algorithm. \href{https://doi.org/10.3390/rs18020214}{\nolinkurl{10.3390/rs18020214}} & Non-spatial decomposition\\
2026 & Gray-code-assisted phase unwrapping method based on phase information partition correction. \href{https://doi.org/10.1364/ao.588065}{\nolinkurl{10.1364/ao.588065}} & Non-spatial decomposition\\
2026 & Interferometric Phase Map Segmentation Algorithm Based on an Adaptive Optimization Tolerance Granularity Space Model. \href{https://doi.org/10.1109/jstars.2026.3709326}{\nolinkurl{10.1109/jstars.2026.3709326}} & No phase unwrapping, or partition without unwrapping\\
2026 & Offset-Supported Phase Unwrapping for X-band InSAR: Evaluation Using the 2024--2025 Fentale--Dofen Dyke Intrusion. \href{https://doi.org/10.2139/ssrn.7117562}{\nolinkurl{10.2139/ssrn.7117562}} & No partition fixed before the local solves\\
2026 & Real-Time Deformation Monitoring and Imaging for Long-Track SAR Based on GPU-Accelerated Parallel Multistream Processing. \href{https://doi.org/10.1109/tim.2026.3654734}{\nolinkurl{10.1109/tim.2026.3654734}} & No phase unwrapping, or partition without unwrapping\\
\end{longtable}}

\end{document}